\documentclass[a4paper,fleqn,usenatbib]{mnras}

\usepackage{newtxtext,newtxmath}

\usepackage[T1]{fontenc}
\usepackage{ae,aecompl}

\usepackage{graphicx}	
\usepackage[caption=false]{subfig}
\usepackage{amsmath}	
\usepackage{amssymb}	
\usepackage{booktabs}
\usepackage{multirow}
\usepackage{gensymb}
\usepackage{threeparttable}
\usepackage{threeparttablex}
\usepackage{todonotes}

\usepackage[separate-uncertainty=true]{siunitx}
\DeclareSIUnit\parsec{pc}
\DeclareSIUnit\solarmass{M\ensuremath{_{\odot}}}
\DeclareSIUnit\year{yr}

\usepackage{longtable}

\usepackage[nonumberlist]{glossaries}

\newcommand{\co}{$^{12}$CO}
\newcommand{\tco}{$^{13}$CO}
\newcommand{\coone}{$^{12}$CO $J$=1-0}
\newcommand{\tcoone}{$^{13}$CO $J$=1-0}
\newcommand{\cotwo}{$^{12}$CO $J$=2-1}

\newcommand{\htwo}{H$_2$}

\newcommand{\alphaco}{$\alpha_{\mathrm{CO}}$}
\newcommand{\alphacou}{$\si{M_{\odot}\ (K\ km\ s^{-1}\ pc^{2})^{-1}}$}
\newcommand{\cothree}{$^{12}$CO $J$=3-2}
\newcommand{\abun}{[$^{12}$CO]/[$^{13}$CO]}

\title[Molecular Gas and Star Formation Properties of Arp 240]{Is this an Early Stage Merger? A Case Study on Molecular Gas and Star Formation Properties of Arp 240 }

\author[He et al.]{
Hao He$^{1}$\thanks{E-mail: heh15@mcmaster.ca (KTS)}, 
C. D. Wilson$^{1}$,
Kazimierz Sliwa, 
Daisuke Iono$^{2,3}$,
and Toshiki Saito$^{2,4,5}$ 
\\
$^{1}$Department of Physics and Astronomy, McMaster University, Hamilton, ON L8S4M1, Canada\\
$^{2}$National Astronomical Observatory of Japan, 2-21-1 Osawa, Mitaka, Tokyo, 181-0015, Japan \\
$^{3}$ SOKENDAI (The Graduate University for Advanced Studies), 2-21-1 Osawa, Mitaka, Tokyo 181-8588, Japan \\
$^{4}$Department of Astronomy, The University of Tokyo, 7-3-1 Hongo, Bunkyo-ku, Tokyo 113-0033, Japan \\
$^{5}$Max-Planck Institute for Astronomy, K¨onigstuhl 17, D69117, Heidelberg, Germany
}

\date{Accepted XXX. Received YYY; in original form ZZZ}

\pubyear{2019}

\begin{document}
\label{firstpage}
\pagerange{\pageref{firstpage}--\pageref{lastpage}}
\maketitle

\begin{abstract}
We present new high resolution $^{12}$CO $J$=1-0, $J$=2-1, and $^{13}$CO $J$=1-0 maps of the early stage merger Arp 240 (NGC5257/8) obtained with the Atacama Large Millimeter/submillimeter Array (ALMA). Simulations in the literature suggest that the merger has just completed its first passage; however, we find that this system has a lower global gas fraction but a higher star formation efficiency compared to typical close galaxy pairs, which suggests that this system may already be in an advanced merger stage. We combine the ALMA data with $^{12}$CO $J$=3-2 observations from the Submillimeter Array and carry out RADEX modeling on several different regions. Both the RADEX modeling and a local thermal equilibrium (LTE) analysis show that the regions are most likely to have a CO-to-H$_2$ conversion factor $\alpha_{\mathrm{CO}}$ close to or perhaps even smaller than the typical value for (ultra-)luminous infrared galaxies. Using 33 GHz data from the Very Large Array to measure the star formation rate, we find that most star forming regions have molecular gas depletion times of less than 100 Myr. 
We calculated the star formation efficiency (SFE) per free-fall time for different regions and find some regions appear to have values greater than 100\%. We find these regions generally show evidence for young massive clusters (YMCs). After exploring various factors, we argue that this is mainly due to the fact that radio continuum emission in those regions is dominated by that from YMCs, which results in an overestimate of the SFE per free-fall time. 
\end{abstract}




\section{Introduction}

In the 1980s, the IRAS satellite was used to identify  a new class of objects with extremely high infrared luminosity called LIRGs and ULIRGs (Ultra/Luminous Infrared Galaxies).  Most of these objects have been identified as merging pairs of galaxies with extreme starburst activity \citep{Sanders_1988, Howell_2010, Elbaz_2011}. As a laboratory for star formation (SF) in extreme environments, many of these mergers have been studied in molecular gas (as traced by CO) at high resolution \citep[e.g.][]{Wilson_2008, Sliwa_2017}. However, most of these studies have focused on intermediate or late-phase mergers, with few observations of early-stage mergers when the two galaxies are still well separated (d$_{proj}>$ 40 kpc). These objects are gaining increasing attention in statistical studies to understand galaxy evolution along the merging sequence. According to many studies \citep[e.g.][]{Violino_2018, Pan_2018}, the average star formation efficiency (SFE=SFR/M$_{H_2}$) is not enhanced significantly as the projected distance decreases, and so the major enhancement \citep[a factor of $\sim$ 2,][]{Ellison_2013} seen in the SFR is attributed  to an increase of the H$_2$ gas mass. However, the large scatter seen in the SFE could hide a dependence on the physical properties of the molecular gas. 
On the other hand, \citet{Moreno_2019} show in their simulation that the increase in the amount of molecular gas during the merging process is only slightly correlated with the SFR enhancement (SFR$_{\text{merger}}$/SFR$_{\text{isolated}}$). Therefore, 
to fully understand the merging process, we need additional resolved studies focused on the molecular gas environment in early mergers. 

Simulations have shown that the inflow of gas caused by tidal interactions will trigger a starburst in the center of the galaxy \citep{Mihos_Hernquist_1996}, which is observed in most cases. However, some mergers, such as the Antennae (NGC4039/39), show an off-nuclear starburst \citep{Schirm_2014,Bemis_2019}. High resolution (1 pc scale) simulations show that compressive turbulence is responsible for this off-center starburst location \citep{Teyssier_2010, Renaud_2014}. Unlike normal spiral galaxies, which have $\Sigma_{\text{mol}} \sim \Sigma_{\text{SFR}}$, the Kennicutt-Schmidt law becomes superlinear in these starburst galaxies with high $\Sigma_{\text{mol}}$ \citep{Daddi_2010}. \cite{Usero_2015} found that while the depletion time $t_{\text{dep}}$= 1/SFE decreases as $\Sigma_{\text{mol}}$ increases, the depletion time for dense gas, which is traced by HCN emission, stays almost constant in the whole range of $\Sigma_{\text{mol}}$. This could be explained by turbulence models \citep{Krumholz_Mckee_2005}. In this model, the turbulence will set the probability density function (PDF) of the clouds. Only the dense part of the cloud will collapse and form stars on the time scale of the cloud free-fall time. In this model, the fraction of gas to form stars can be represented by the SFE per free-fall time, which is 
\begin{equation}
\epsilon_{\text{ff}}=\frac{t_{\text{ff}}}{t_{\text{dep}}}
\end{equation} 
\cite{Krumholz_2012} argued that $\epsilon_{\text{ff}}$ should be 2\% on all scales with only a factor of 3 scatter. However, \cite{Lee_2016} calculate $\epsilon_{\text{ff}}$ for local Galactic clouds and find that the scatter in this quantity is more like a factor of 8 with a maximum value of several 10\%. \cite{Semenov_2016} found that the distribution of cloud virial parameters is broad enough to account for the observed scatter of $ \epsilon_{\text{ff}}$ for an individual galaxy.
Overall, $ \epsilon_{\text{ff}}$ still seems to be highly dependent on the environment. Therefore, we need to explore the molecular gas properties in detail to understand star formation in these regions. 

Since molecular gas does not produce \htwo\ emission lines under normal physical conditions, we generally use \coone\ emission to trace the amount of molecular gas. This will introduce the conversion factor \alphaco\ between the surface density of the molecular gas and the \coone\ intensity. \alphaco\ in the Milky Way is found to be 4.3 $\si{M_{\odot}\ pc^{-2}\ (K\ km\ s^{-1})^{-1}} $ \citep{Bolatto_2013}. In LIRGs and ULIRGs, the typical conversion factor is found to be smaller, with a typical value of 1.1 $\si{M_{\odot}\ pc^{-2}\ (K\ km\ s^{-1})^{-1}}$ \citep[][including helium]{Downes_Solomon_1998}. \citet{Herrero-Illana_2019} also find a similar value of 1.8 $\si{M_{\odot}\ pc^{-2}\ (K\ km\ s^{-1})^{-1}}$ for these U/LIRGs assuming a fixed gas-to-dust mass ratio. \citet{Narayanan_2011} explored the difference in conversion factors between mergers and normal disk galaxies in simulations. They found \alphaco\ will decrease as merging starts and then eventually come back to the Milky Way value as the merging event finishes. They also found that the combination of increasing the velocity dispersion and the temperature of giant molecular clouds (GMCs) is what causes the conversion factor to be lower than the Milky Way value by a factor of 2 to 10. In addition, \citet{Renaud_2019} show in their simulation that \alphaco\ correlates more tightly with depletion time among different regions in their simulated mergers. They conclude that the change in \alphaco\ is also mainly driven by the energy feedback and velocity dispersion. However, \citet{Papadopoulos_2012} point out in their data that the major reason for small the \alphaco\ in LIRGs and ULIRGs is the large velocity dispersion. The cause of the low conversion factor in these systems is still under debate.  

One of the tools to constrain the molecular gas properties as well as the conversion factor is RADEX modeling \citep{vandertak_2007}. To perform this kind of analysis, we need multiple CO lines. This kind of analysis is often performed over entire galaxies with large beams \citep[e.g. ][]{Kamenetzky_2016}. With the help of ALMA, we can perform this analysis in resolved regions of individual galaxies \citep[e.g. ][]{Sliwa_2012, Saito_2017, Sliwa_2017, Sliwa_Downes_2017}. These studies enable us to make a direct comparison between mergers in different stages. \citet{Sliwa_2017} compared the early stage merger Arp 55 with the late stage merger NGC 2623. They found that for an early merger like Arp 55, the conversion factor is still well below the Milky Way value. They argue that Arp 55 is not an early enough merger to catch the transition of \alphaco\ from the Milky Way value to the ULIRG value. A study of an even earlier merger will help us to explore this problem. 

In this paper, we study Arp 240, which is another LIRG at an early merging stage. Arp 240 is composed two massive spiral galaxies, NGC 5257 and NGC 5258. The two galaxies are intertwined in HI \citep{Iono_2005} but well separated in optical images. The projected distance of the two galaxies is about 40 kpc, which is larger than the separation of Arp 55 ($\sim$9 kpc). Simulations \citep{Privon_2013} show the two galaxies are in the early stage of the merging process and have just been through first passage. 
It is worth noting that Arp 55, which is also identified to be around its first passage by visual classification \citep{Haan_2013, Stierwalt_2013}, is probably at an earlier stage than Arp 240 due to its smaller separation. However, since merging stage depends on various quantities, it is difficult to be certain. More reliable dynamic modeling would help to define the merging stage of Arp 240 more precisely. Basic information on Arp 240 is listed in Table \ref{tab:basic}. In studying this system, we hope to get a better understanding of the star formation activity and gas physical properties in an early merger system. 

This paper is organized as follows. In Section 2, we describe basic information on the observations and how we processed the data. In Section 3, we report our measurements of several quantities, such as gas mass, line ratio, and SFR in different regions. In Section 4, we carry out a RADEX analysis to explore the gas physical properties and conversion factors in different regions. In Section 5, we use \cotwo\ data and 33 GHz continuum data to explore the relationship between molecular gas and SFR under the framework of the turbulence model.

\begin{table*}
	\centering
	\caption{Basic properties of NGC 5257 and NGC 5258}
	\begin{threeparttable}
		\begin{tabular}{lccc}
			\hline
			& NGC 5257                                        & NGC 5258                                        & \# References \\ \hline
			Coordinates (J2000) $^a$                 & RA$=13^h39^m52.91^s$                            & RA$=13^h39^m57.70^s$                            & ...           \\
			& Dec=+00\degree50\arcmin24.5\arcsec & Dec=+00\degree49\arcmin51.1\arcsec & ...           \\
			Morphological Type $^b$                  & SAB(s)b pec                             & SA(s)b:pec                            & ...           \\
			Redshift $^b$                            & 0.022676                                        & 0.022539                                        & ...           \\
			Luminosity Distance (Mpc)                & 98.0                                            & 97.4                                            & 1             \\
			$L_{\text{H} \alpha}\ (10^6\ \si{L_{\odot}}$)        & 5                                               & 6                                               & 2             \\
			$L_{\text{TIR}}$ ($10^{11}\ \si{M_{\odot}}$) $^c$    & 1.3                                             & 1.5                                             & This work    \\
			HI mass ($10^{10}\ \si{M_{\odot}}$)            & 1.2                                             & 0.98                                            & 3             \\
			H$_2$ mass ($10^{9}\ \si{M_{\odot}}$) $^d$   & 4.6                                            & 7.2                                             & This work    \\
			Stellar Mass ($10^{10}\ \si{M_{\odot}}$) $^e$ & 9.4                                             & 10.5                                            & This work    \\
			SFR ($\si{M_{\odot}}$ yr$^{-1}$) $^f$         & 27.8                                            & 24.9                                            & 3,4           \\ \hline
		\end{tabular}
		\begin{tablenotes}
			\item \textbf{Notes.} $^{(a)}$HyperLEDA. $^{(b)}$NED. $^{(c)}$From the combination of 24 \textmu m image from \textit{Spitzer} and 70 \textmu m image from \textit{Herschel}. $^{(d)}$From the \coone\ ALMA observations assuming a 	ULIRG conversion factor of 1.1 \alphacou. $^{(e)}$From the 3.6 \textmu m and 4.5 \textmu m \textit{Spitzer} data. $^{(f)}$From L$_{1.4 \textrm{GHz}}$. 
			\item \textbf{References.} (1) \citet{Mould_2000}; (2) \citet{Sofue_1993}; (3) \citet{Iono_2005}; (4) \citet{Yun_2001}
		\end{tablenotes}
	\end{threeparttable}
	\label{tab:basic}
\end{table*}

\section{Observations and Data Reduction}
We use multiple CO lines (\coone, $J$=2-1 and \tcoone ) from the Atacama Large Millimeter/Submillimeter Array (ALMA) to determine the physical properties of the gas in Arp 240. To further constrain the properties of the gas, we add \cothree\ data from the Submilimeter Array (SMA). The SFR is traced by infrared data and radio continuum. We use the 24 \textmu m map from the \textit{Spitzer} Space Telescope, the 70 \textmu m map from the \textit{Herschel} Space Telescope, and the 33 GHz continuum map from the Very Large Array (VLA) to trace the SFR. 

\subsection{ALMA Data}
\label{sec:maths} 

\begin{table*} 
	\centering
	\caption{Summary of ALMA molecular line observations}
	\label{tab:observation}
	\begin{tabular}{lcccccc} 
		\hline
		Molecular  & RMS noise  & Beam (") &  Field & Channel  & Observed  \\
		line & (mJy beam$^{-1}$) & & & Width (km s$^{-1}$) & Frequency (GHz) \\
		\hline
		\coone\ & $1.6 $ & 2.0 $\times$ 1.6 & 12m+7m array   & 10 & 112.73\\
		\cotwo\ & $3.0 $ & 1.0 $\times$ 0.5 & 12m+7m array   & 10 & 225.46\\
		\tcoone\ & $0.64 $ & 2.1 $\times$ 1.6 & 12m array  & 20 & 107.78\\
		\hline
	\end{tabular}
\end{table*}

\begin{figure*}
	\centering
	\subfloat{\includegraphics[width=0.35\linewidth]{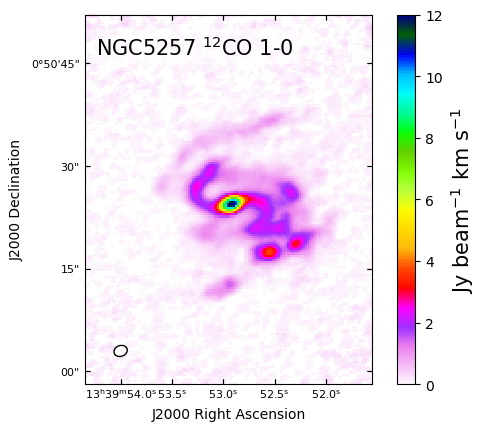}} 
	\hspace{1.0cm}
	\subfloat{\includegraphics[width=0.35\linewidth]{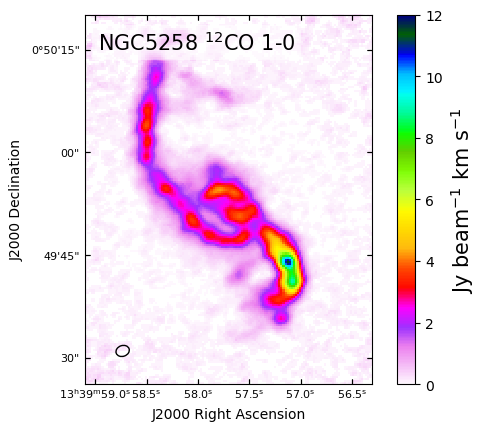}} \\
	\subfloat{\includegraphics[width=0.35\linewidth]{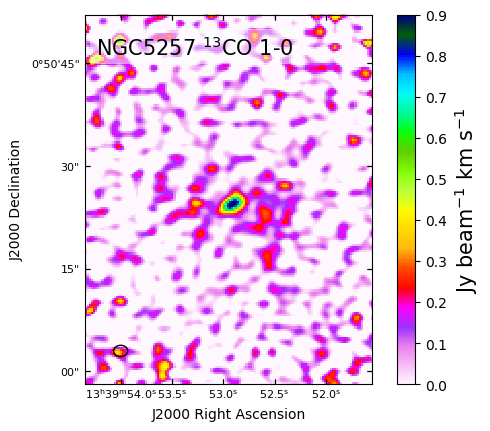}} 
	\hspace{1.0cm}
	\subfloat{\includegraphics[width=0.35\linewidth]{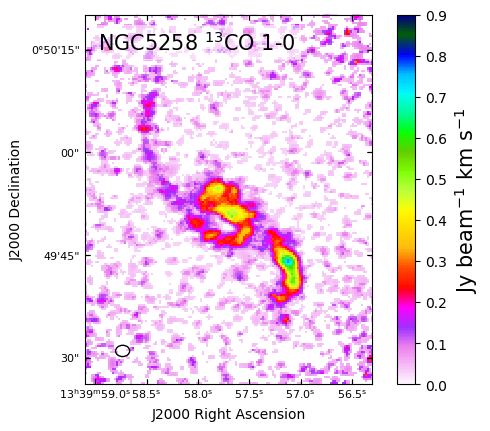}} \\
	\subfloat{\includegraphics[width=0.35 \linewidth]{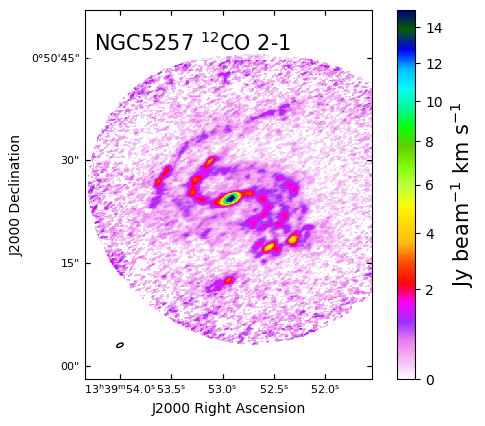}} 
	\hspace{1.0cm}
	\subfloat{\includegraphics[width=0.35\linewidth]{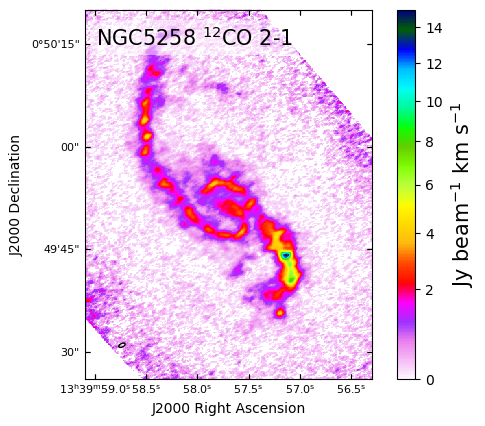}}
	\caption[Integrated intensity map of \coone\, \tcoone\ and \cotwo\ for NGC 5257 and NGC 5258]{From top to bottom are the integrated intensity maps of \coone, \tcoone\ and \cotwo\ for NGC 5257 (left) and NGC 5258 (right). The black ellipse in each plot indicates the size of the beam. The size of the beams are 2.0\arcsec $\times$ 1.6\arcsec , 2.1\arcsec $\times$ 1.6\arcsec\ and 1.0\arcsec $\times$ 0.5\arcsec\ for \coone, \tcoone\ and \cotwo\ respectively.}
	\label{fig:mom0}
\end{figure*}

\begin{figure*}
	\centering
	\subfloat{\includegraphics[width=0.35\linewidth]{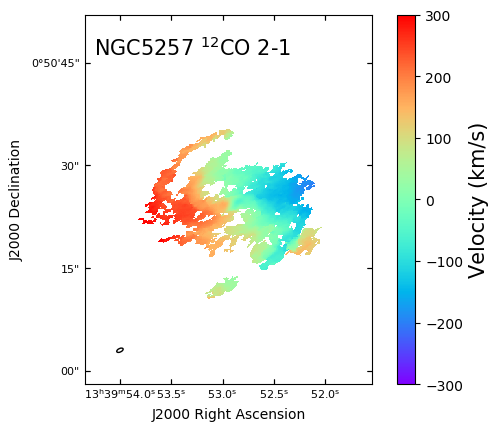}}%
	\hspace{1.0cm} 
	\subfloat{\includegraphics[width=0.35\linewidth]{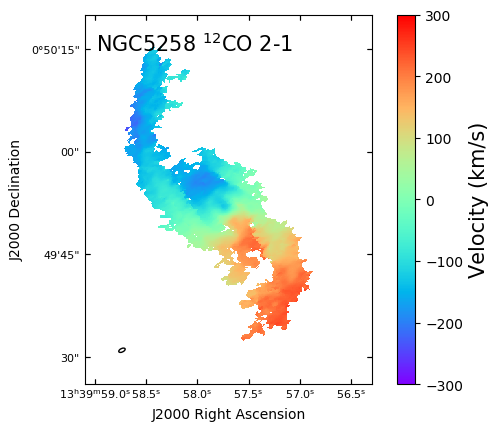}} \\
	\subfloat{\includegraphics[width=0.35\linewidth]{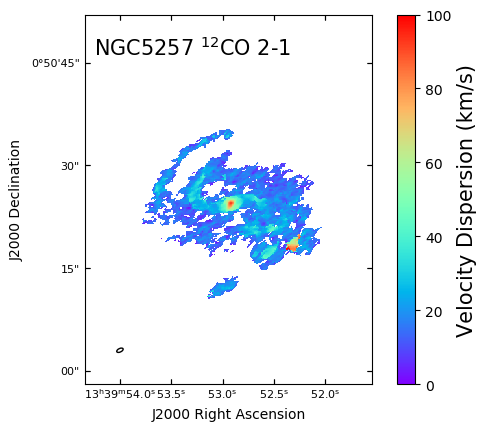}} 
	\hspace{1.0cm}
	\subfloat{\includegraphics[width=0.35\linewidth]{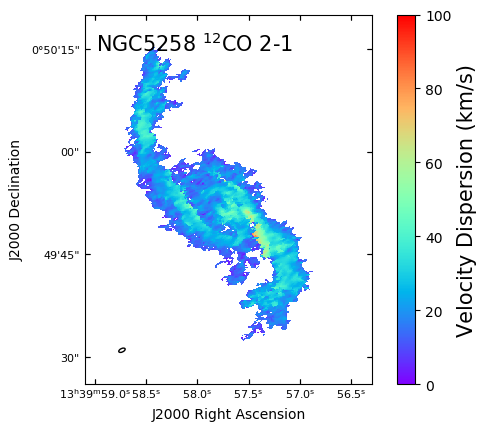}}
	\caption [Velocity field (left) and velocity dispersion (right) map for NGC 5257 and NGC 5258]{Velocity field and dispersion map of NGC 5257 (left) and NGC 5258 (right) derived from the \cotwo\ data cube. The beam size for the \cotwo\ cube is 1.0\arcsec $\times$ 0.5\arcsec. }
	\label{fig:vel}
\end{figure*}

The data for Arp 240 was acquired from project 2015.1.00804.S (PI: Kazimierz Sliwa). Single pointings were used for the Band 3 data. For the Band 6 data, four pointings were used for NGC 5257 while 9 pointings were used for NGC 5258. The resolution of Band 3 is about 2 arcsec (1 kpc) while the resolution of Band 6 is about 1 arcsec (0.5 kpc). Band 3 covers \coone\ (112 GHz) and \tcoone\ (108 GHz) and Band 6 covers \cotwo\ (223 GHz). We also detected CN $J$=1-0 (111 GHz) and CS $J$=2-1 (95.8 GHz) in Band 3. The total usable bandwidth for each of the spectral windows was 1875 MHz for the 12m array and 2000 MHz for the 7m array. The width of a single channel is 1.953 MHz. 

The original reduction scripts were used to calibrate the raw data using CASA version 4.5.0, which is the version of CASA when the data was taken. We use CASA 5.1.1 to select line free channels and subtract the continuum from the line cubes using \texttt{uvcontsub} command. All imaging steps were carried out in CASA 5.4.0. We used the command \texttt{tclean} and set the channel width to be 10 km s$^{-1}$ for \coone\ and \cotwo\ and 20 km s$^{-1}$ for \tcoone\ to achieve better sensitivity. The total velocity range is set to be 500 km s$^{-1}$. For cleaning, we set the threshold to be 2 times the RMS noise. We use the auto multithreshold option in the \texttt{tclean} command to identify clean regions automatically. There are four key parameters: noisethreshold, sidelobethreshold, lownoisethreshold and negativethreshold. We generally use the default setting except for \cotwo. We found strong sidelobes at the edges of the \cotwo\ map for NGC 5257 and therefore we set the sidelobethreshold to be 4.0 instead of the default 3.0.

After imaging, we created moment 0 maps (Fig. \ref{fig:mom0}) using the CASA command \texttt{immoments} with threshold of 2 RMS for each image cube. We also created moment 1 and moment 2 maps (shown in Fig. \ref{fig:vel}) following the procedure used by \citet{Sun_2018}. The advantage of this method compared to the simple threshold method is that it will only pick out signals spanning more than 2 consecutive velocity channels, which excludes noisy pixels with inaccurate velocity measurements. A more detailed description of their algorithm can be found in their Section 3.2. This algorithm was implemented in a Python script\footnote{\url{https://github.com/astrojysun/Sun_Astro_Tools/blob/master/sun_astro_tools/spectralcube.py}} and graciously made available for use. 

Before we make ratio maps, we need to make sure all the data have a similar beam size and UV coverage. We cut the inner uvrange of the \coone\ and \tcoone\ data at 6 k$\lambda$ to match the large scale. We also uvtaper the \cotwo\ data in \texttt{tclean} in CASA with a Gaussian beam with size of 1.65\arcsec $\times$ 1.44\arcsec\ in image space to match the small scale. After re-making the image cube, we smoothed all the image cubes to have a beam size of 2.186\arcsec\ $\times$ 1.896\arcsec. After cutting the uvrange, the largest angular scale (LAS) is about 32 arcsec. 


\subsection{SMA Data}

\begin{figure*}
	\centering
	\subfloat{\includegraphics[width=0.4\linewidth]{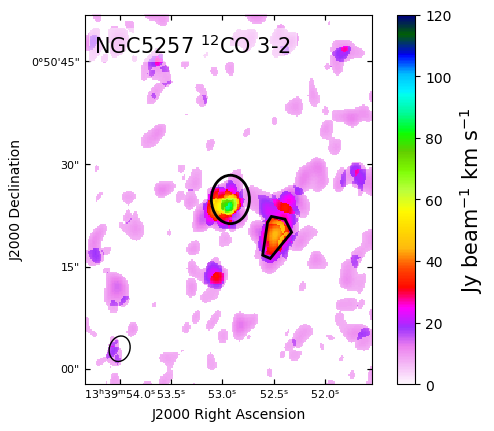}}%
	\hspace{1.0cm}
	\subfloat{\includegraphics[width=0.4\linewidth]{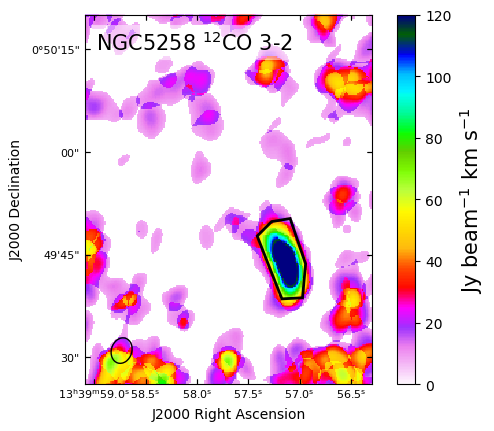}}%
	\caption[SMA moment 0 map of \cothree\ for NGC 5257 and NGC 5258 ]{The SMA moment 0 map of \cothree\ for NGC 5257 (left) and NGC 5258 (right). Regions encircled by black apertures are used in the RADEX modeling. }
	\label{fig:SMA}
\end{figure*}

Arp 240 was observed with the SMA in \cotwo\ and \cothree\ using the compact array configuration. The \cothree\ map has a beam size of 3.79\arcsec\ $\times$ 2.8\arcsec. The data processing is described in \citet{Wilson_2008}. The moment 0 map is made from a cube with channel width of 40 km s$^{-1}$ with a threshold of 2 RMS (127 mJy/beam). In this study, we remake the moment 0 map (Fig. \ref{fig:SMA}) using the CASA \texttt{immoments} command to correct an error\footnote{This error only affected the moment 0 map and not any of the fluxes given in \citet{Wilson_2008}.} in the velocity range used in producing the moment 0 map shown in \citet{Wilson_2008}.


\subsection{Infrared Data}

We obtained the \textit{Spitzer} 3.6 \textmu m and 4.5 \textmu m data from the archive. The \textit{Spitzer} 3.6 \textmu m and 4.5 \textmu m data is from project 70038 (PI:Sanders) with a resolution of roughly 2 arcsec. We can use the data to calculate the stellar mass in the galaxies. The relevant equation is given in \citet{Eskew_2012}
\begin{equation}
M_{\star}=10^{5.65}F_{3.6}^{2.85}F_{4.5}^{-1.85} (\frac{D}{0.05})^2
\end{equation}
where $M_{\star}$ is the stellar mass in $\si{M_{\odot}}$. $F_{3.6}$ and $F_{4.5}$ are fluxes in Jy and $D$ is the distance of the source in Mpc. This equation assumes a Salpeter Initial Mass Function (IMF). To convert to the Kroupa IMF, we multiply the result by 0.7. 

\subsection{Radio Continuum}


Arp 240 was observed with the VLA in 33 GHz continuum. The detailed description of the data is in \citet{Linden_2019}. The resolution of the data is about 0.5 arcsec and the sensitivity is about $9.8 \times 10^{-6}$ Jy beam$^{-1}$. The relation between the radio continuum and the SFR has been calibrated \citep{Murphy_2011} as
\begin{equation}
\begin{split}
\mathrm{SFR}=&10^{-27}[2.18(\frac{T_e}{10^4\ K})^{0.45}(\frac{\nu}{\si{GHz}})^{-0.1}+15.1 \\
&\times (\frac{\nu}{\si{GHz}})^{-\alpha^{NT}}]^{-1}(\frac{L_{\nu}}{\si{erg\ s^{-1}\ Hz^{-1}}})
\end{split}
\end{equation}
where SFR is in solar masses per year, $T_e$ is the electron temperature in Kelvin, $\nu$ is the observed frequency in GHz and $L_{\nu}$ is the luminosity of the source in $\si{erg\ s^{-1}\ Hz^{-1}}$. $\alpha^{NT}$ is the spectral index for synchrotron emission. We assume $\alpha^{NT}=0.85$ \citep{Murphy_2012} and $T_e=10^4\ \si{K}$.

\section{Measurements}

\subsection{Gas Mass}

The \coone\ line is a commonly used tracer for molecular gas mass in galaxies. The equation to calculate the gas mass from the CO luminosity is
\begin{equation}
M_{\text{mol}}= \alpha_{\text{CO}} \times L_{\text{CO}}(1-0)
\end{equation}
where $M_{\text{mol}}$ is the molecular gas mass in $\si{M_{\odot}}$, $L_{\text{CO}} (1-0)$ is the \coone\ luminosity in $\si{K\ km\ s^{-1}\ pc^2}$ and \alphaco\ is the CO-to-\htwo\ conversion factor in \alphacou. 
$L_{\text{CO}}$ is calculated as \citep{Bolatto_2013}
\begin{equation}
L_{\text{CO}}(1-0)= 2453\  S_{\text{CO}} \Delta v D_L^2/(1+z)
\end{equation}
where $S_{\text{CO}} \Delta v$ is the integrated flux in Jy kms$^{-1}$, $D_L$ is the luminosity distance to the source in Mpc and z is the redshift of the source. The conversion factor \alphaco\ varies among different types of galaxies. In order to roughly quantify the conversion factor of the molecular gas, we adopt the recipe in \citet{Violino_2018}. In their paper, they calculate the conversion factor as 
\begin{equation}
\alpha_{\text{CO}}= (1-f_{\text{SB}}) \times \alpha_{\text{CO, MS}} + f_{\text{SB}} \times \alpha_{\text{CO, SB}}
\end{equation}
where $f_{\text{SB}}$ is the probability for a galaxy to be a starburst galaxy. This probability is determined by the offset of the specific star formation rate (sSFR) from the star-forming main sequence \citep{Sargent_2014}. For the expected sSFR of the main sequence versus a function of stellar mass, we adopted the equation from \citet{Catinella_2018}, 
\begin{equation}
\log \mathrm{sSFR}_{MS}= -0.344(\log M_{\star}-9)-9.822
\end{equation}
with an uncertainty of
\begin{equation}
\sigma_{\text{MS}}=0.088(\log M_{\star}-9)+0.188
\end{equation}
Then we calculate the ratio between the actual sSFR and the expected 
sSFR from the main sequence fitting relation. The ratio is 9.37 for NGC 5257 and 7.8 for NGC 5258. This corresponds to $f_{\text{SB}} \approx 1$. This analysis suggests that we should adopt the ULIRG conversion factor for both galaxies. 
We adopted a circular aperture with radius equal to the Petrosian radius $R_p$ to measure the global gas mass. $R_p$ is 17.53 arcsec for NGC 5257 and 24.86 arcsec for NGC 5258 from the SDSS DR12 catalog\footnote{ \url{http://skyserver.sdss.org/dr12/en/tools/explore/}}.
The gas mass is about $4.6 \times 10^{9}\  \si{M_{\odot}}$ for NGC 5257 and $7.2 \times 10^{9}\  \si{M_{\odot}}$ for NGC 5258 adopting the ULIRG conversion factor.

By comparing the moment 0 maps (Fig. \ref{fig:mom0}) from all three lines, we can see the morphologies of the different line tracers are almost the same. Therefore, we use the original \coone\ image (before applying the uvrange cut) to study the molecular gas distribution among different regions of the galaxies. For NGC 5257, the gas is clearly concentrated in the center. To quantitatively learn about the gas concentration degree, we calculated $\Sigma_{\text{mol, 500 pc}}/ \Sigma_{\text{mol}, R25}$, which is the ratio between the gas surface density in the central 500 pc and within isophotal radius R25, which is 53.3 arcsec for NGC 5257 \citep{Fuentes-Carrera_2019}.  Due to the limited sensitivity, we do not detect molecular gas out to the isophotal radius
and there are lot of blank pixels in the outer regions which have values below the 2 RMS threshold cut. In fact, some regions are even out of our field of view. We treated these blank pixels within the isophotal radius in two ways.
The first is to assume all the blank pixels have an intensity equal to zero. The second is to assume all the blank pixels have an intensity equal to the noise, which defines the lower limit of the ratio. We calculated the concentration degree of NGC 5257 to be $72 \sim 95$.  \citet{Sakamoto_1999} compared the gas concentration degree of barred galaxies and unbarred galaxies. They found the concentration degree is $100.2 \pm 69.8$ (the error is the standard deviation for 10 objects) for barred galaxies and $24.9 \pm 18.5$ for unbarred galaxies. This clearly suggests that NGC 5257 is fairly gas concentrated in the center. We also put a 500 pc aperture in the center of NGC 5258 and calculated the concentration degree in the center, which is about $11 \sim 13$. This value is among the typical values for normal disk galaxies. For NGC 5258, the gas is concentrated in the south spiral arm instead of the center.

\begin{figure*}
	\centering 
	\subfloat{\includegraphics[width=0.45\linewidth]{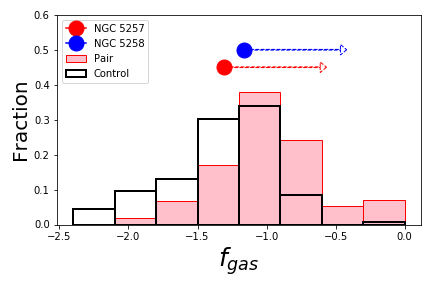}}
	\hspace{1.0 cm}
	\subfloat{\includegraphics[width=0.45\linewidth]{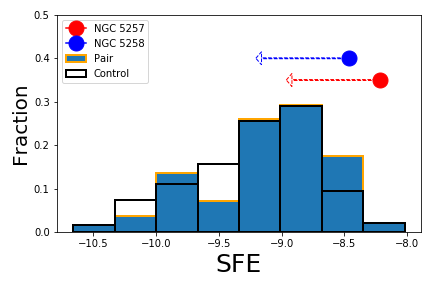}}
	\caption[Global gas fraction of NGC 5257 and NGC 5258]{(Left) The gas fraction histogram of individual galaxies in the pairs and control samples from \citet{Pan_2018}. The global gas fractions of NGC 5257 and NGC 5258 are overlaid on the histogram. The upper and lower limits use the Miky Way and ULIRG conversion factor respectively. (Right) The SFEs of NGC 5257 and NGC 5258 overplotted on the SFE histogram of individual galaxies in the pairs and control sample from \citet{Pan_2018}. The lower and upper limits correspond to the Milky Way and ULIRG \alphaco\ respectively.  }
	\label{fig:histogram}
\end{figure*}

We also calculated the global molecular gas to stellar mass fraction and SFE in both galaxies. The global stellar mass is calculated with \textit{Spitzer} 3.6 \textmu m and 4.5 \textmu m images (see Section 2.3) with the same aperture that is used to measure the global gas mass. The global SFR is taken from Table \ref{tab:basic}.  
From Figure \ref{fig:histogram}, we can see that the gas fraction is similar to normal disk galaxies. On the contrary, the SFE of both galaxies is at the higher end of the distribution. By comparing NGC 5257 and NGC 5258, we can see NGC 5258 has higher gas fraction and lower SFE, which might suggest NGC 5258 has more gas yet to be converted to stars and therefore is in a younger star forming stage.

\subsection{Line Ratio}

\begin{figure*}
	\centering 
	\subfloat{\includegraphics[width=0.4\linewidth]{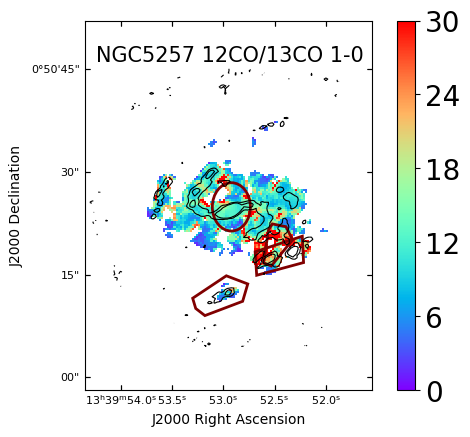}}%
	\hspace{1.0cm}
	\subfloat{\includegraphics[width=0.4\linewidth]{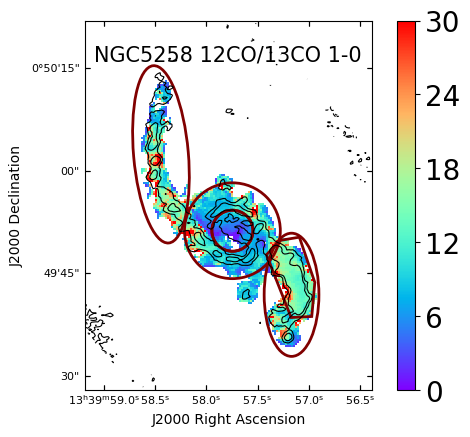}} \\
	\subfloat{\includegraphics[width=0.4\linewidth]{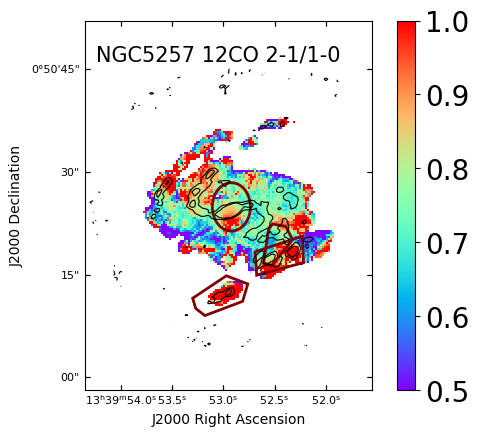}} 
	\hspace{1.0cm}
	\subfloat{\includegraphics[width=0.4\linewidth]{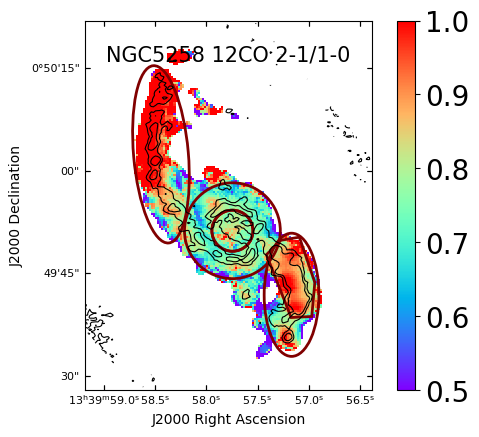}}
	\caption[Brightness temperature ratio maps of different molecular lines for NGC 5257 and NGC 5258]{Brightness temperature ratio maps of different molecular lines for NGC 5257 (left) and NGC 5258 (right). The left column is the \co/\tcoone\ ratio map and the right column is the \cotwo/1-0 ratio map. The \cotwo\ moment 0 map in Figure \ref{fig:mom0} is overlaid as contours for reference of the aperture location. The contour levels are 1.1 and 2.2 Jy beam$^{-1}$ km s$^{-1}$.
	The maroon regions are the apertures we used to measure the flux ratio (see text for details). The central ellipse and south-west pentagon in NGC 5257 and the south-west polygon in NGC 5258 are also used to measure the flux for RADEX modeling, as shown in Fig. \ref{fig:SMA}.}
	
	\label{fig:ratio}
\end{figure*}

\begin{table}
	\label{tab: ratio}
	\centering
	\caption{Line ratios measured in different regions of NGC 5257 and NGC 5258. }
	\renewcommand{\arraystretch}{1.2}
	\small\addtolength{\tabcolsep}{-2pt}
	\begin{threeparttable}
	\begin{tabular}{llcc}
		\hline
		\multicolumn{2}{l}{}                            & \co/\tco   & \co  \\
		\multicolumn{2}{l}{} & $J$ =1-0 & $J$ = 2-1/1-0   \\ \hline
		NGC 5257             & Center (RADEX)$^a$                   & $13.7 \pm 1.3$ & $0.83 \pm 0.06$ \\
		 & South West               & $23 \pm 4$ & $0.88 \pm 0.07$ \\
		& South West (RADEX)$^a$ & $ 23 \pm 4$ & $ 0.78 \pm 0.06$ \\
		 & South         & $ 8.4 \pm 3.4 $ & $ 1.2 \pm 0.2 $  \\
		 & Global                    & $ 13.1 \pm 1.0$ & $ 0.80 \pm 0.06$ \\
		NGC 5258             & North Arm                & $13.5 \pm 1.2$  & $0.96 \pm 0.08$  \\
		& South Arm                & $13.8 \pm 1.0$  & $0.88 \pm 0.07$  \\
		& South Arm (RADEX)$^a$        & $14.4 \pm 1.1$  & $0.91 \pm 0.07$  \\
		& Center                   & $5.6 \pm 0.5$   & $0.80 \pm 0.07$  \\
		& Ring around Center       & $9.7 \pm 0.8$   & $ 0.77 \pm 0.06$ \\
		& Global                    & $10.9 \pm 0.8$  & $ 0.85 \pm 0.06$ \\ \hline
	\end{tabular}
	\end{threeparttable}
	\begin{tablenotes}
		\item \textbf{Note:} The apertures are overlaid on line ratio images in Fig. \ref{fig:ratio}. 
		\item a. These measurements are used later for RADEX modeling.   
	\end{tablenotes}
\end{table}

We made brightness temperature ratio maps for the different molecular lines to see how the molecular gas properties vary among different regions. We made \co/\tcoone\ ratio maps and \cotwo/1-0 ratio maps. For making the \co/\tcoone\ ratio map, we noticed that the \tcoone\ cube has fewer detected regions in the cubes. To make sure the ratio map only contains pixels with signal detected in both image cubes, we apply different threshold cuts while making moment 0 maps of the two cubes. 
We applied a 2 $\times$ RMS cut for both the \tcoone\ and \coone\ moment 0 maps in order to calculate the typical flux ratio across each galaxy. We then applied a 2 $\times$ the flux ratio $\times$ \tcoone\ RMS cut to the \coone\ cube to make the moment 0 map.
The final step is to do primary beam correction for both moment 0 maps and calculate the ratio map. For the \cotwo/1-0 ratio map, we applied the same procedure to calculate the ratio maps. Since the 2 RMS cut for \coone\ cube is a low threshold, the ratio map will contain a lot of noisy pixels. Therefore, we adopt a S/N cut of 5 instead of 2 for the \coone\ cube.  

Both ratio maps are shown in Figure \ref{fig:ratio}. For NGC 5257, we can see both ratios are generally uniform across the disk. We therefore only drew apertures around regions of interest. The central circle and south-west pentagon shown in the ratio maps encircle regions with \cothree\ detection, as shown in Figure \ref{fig:SMA}. These two apertures are later used to measure fluxes for RADEX modeling. The other south-west quadrilateral encloses a region with emission peaks in both the \coone\ and \cotwo\ maps (Fig. \ref{fig:mom0}). We can see that this region also has high \co/\tcoone\ and \cotwo/1-0 ratios. The south polygon encircles a region with high \cotwo/1-0 ratio, which also corresponds to the south emission peak in 33 GHz continuum (Fig. \ref{fig:33GHz}). For NGC 5258, we can see that both ratios vary among different morphological regions. We therefore divide the galaxy into four regions of center, ring around the center, north arm and south arm. The south-arm polygon corresponds to a region with \cothree\ detection and will be later used for flux measurement for RADEX modeling. Ratios measured in the different apertures are shown in Table \ref{tab: ratio}.

Both galaxies have a global \co/\tcoone\ flux ratio around 10, which is typical for normal spiral galaxies \citep{Cormier_2018}. For NGC 5257, the majority of the disk has a \co/\tcoone\ ratio around 13 except for the south-west regions and the south isolated gas clump. The extremely high \co/\tcoone\ ratio in the two south-west regions could be caused by a high \abun\ abundance ratio, which may suggest the inflow of fresh molecular gas from the outer regions or HI gas converting to molecular gas. The south isolated gas clump has a low \co/\tcoone\ ratio but with a large uncertainty. As we can see from the \tcoone\ map (Fig. \ref{fig:mom0}), there is not much detected in this region. For NGC 5258, different regions have very different ratio values. The central region particularly has low ratio values compared to normal spiral galaxies. This suggests that \tcoone\ in this region is moderately optically thick. Since the central region also has low \coone\ intensity, the low \co/\tcoone\ ratio suggests a low \abun\ abundance ratio there. 

The global \cotwo/1-0 ratio for both galaxies is about 0.8, which is larger than the typical ratio of 0.7 for normal spiral galaxies \citep[references in][]{Sun_2018}. The ratio varies less across different regions for both galaxies. For NGC 5257, the south region has a ratio value above 1.0. This region is identified to have a young ($\sim$ 3.3 Myr), massive ($\sim 10^7\ \si{M_{\odot}}$) star cluster \citep{Linden_2017}. \citet{Smith_2014} also show an extremely bright X-ray source which might correspond to the star cluster. In this case, the high \cotwo/1-0 ratio might be caused by an unusual heating from X-ray emission. A similarly high ratio is seen in the north spiral arm of NGC 5258. However, there is no X-ray emission detected there. The high \cotwo/1-0 ratio may suggest gas in those regions is not in local thermal equilibrium (LTE). 

\subsection{Star Formation Rates}

\begin{table}
	\centering
	\caption{Depletion time of different regions in NGC 5257 and NGC 5258 }
	\captionsetup{justification=centering} 
	\label{tab:tdep}
	\centering
	\begin{threeparttable}
	\begin{tabular}{lcccc}
	\hline
	Galaxy  & Region    & $\Sigma_{\text{SFR}}$$^{a}$ & $\Sigma_{\text{mol}}$$^{b}$ & $t_{\text{dep}}$  \\
	& &  $ \si{(M_{\odot}\ kpc^{-2}\ yr^{-1})}$ & $\si{(M_{\odot}\ pc^{-2})}$ & ($10^8$ yr) \\
	\hline
	NGC5257 & center    & 2.5                                       & 550                               & 2.16                           \\
	& arm       & 1.1                                       & 86                                & 0.77                          \\
	& west & 1.1 & 45 & 0.42 \\
	& south     & 3.0                                        & 79                              & 0.27                          \\
	NGC5258 & south arm & 2.4                                       & 338                               & 1.5                          \\ \hline
\end{tabular}
	\end{threeparttable}
	\begin{tablenotes}
		\item \textbf{Note:} The different regions are identified in Fig. \ref{fig:33GHz}. 
		\item a. $\Sigma_{\text{SFR}}$ is calculated using the 33 GHz image 
		\item b. $\Sigma_{\text{mol}}$ is calculated using \cotwo\ assuming \cotwo/1-0 ratio of 0.8 and typical ULIRG conversion factor 
	\end{tablenotes}
\end{table}

\begin{figure*}
	\centering
	\subfloat{\includegraphics[width=0.45\linewidth]{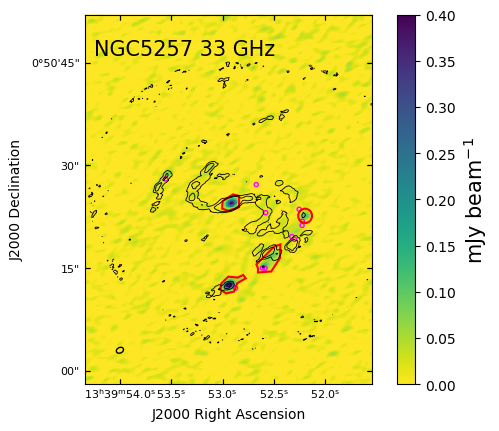}}%
	\hspace{1.0cm}
	\subfloat{\includegraphics[width=0.45 \linewidth]{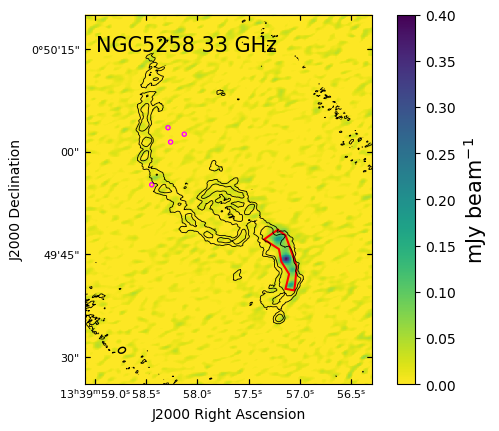}} 
	\caption [33 GHz continuum image for NGC 5257 and NGC 5258 ]{33 GHz continuum image for NGC 5257 (left) and NGC 5258 (right) smoothed to the beam size of 1.1\arcsec $\times$ 0.8\arcsec overlaid by the contour \cotwo\ moment 0 map in Fig. \ref{fig:mom0}. The level of contours are 1.1 and 2.2 Jy beam$^{-1}$ km s$^{-1}$. We divide both galaxies into different regions based on the 33 GHz intensity. NGC 5257 is divided into center, south continuum source, south-west arm and west region. NGC 5258 only has a strong detection in the south spiral arm. The magenta circles represent identified young (age < 10 Myr) massive (mass > $10^6\ \si{M_{\odot}}$) star clusters in \citet{Linden_2017} using HST photometry.}
	\label{fig:33GHz}
\end{figure*}

\citet{Murphy_2011, Murphy_2012, Murphy_2018} use 33 GHz continuum as a SFR tracer. The radio continuum mainly comes from 2 sources, free-free and synchrotron. We use the composite equation from their paper to calculate the SFR, as shown in section 2.4. The 33 GHz continuum maps are shown in Fig. \ref{fig:33GHz}.

We also draw polygon apertures around different SF regions to measure $\Sigma_{\text{SFR}}$, which is shown in Table \ref{tab:tdep}. These polygon regions drawn by eye roughly correspond to regions with signal-to-noise (S/N) value greater than 4.0. The SFR enclosed in the center of NGC5257 makes up only 7\% of the global SFR of NGC 5257, which is similar to the south continuum source in NGC 5257. For NGC 5258, the only radio detected SF region contains about 30\% of the global SFR of this galaxy. We also calculated the surface density of molecular gas in these polygon apertures, as shown in Table \ref{tab:tdep}. Based on these values, we can calculate the average depletion time in different regions, which is also shown in Table \ref{tab:tdep}. We can clearly see the depletion times vary among different regions. In particular, we can see the off-center star forming regions in NGC 5257, such as the south continuum bright source, have short depletion times that are comparable to typical ULIRG values \citep[$\sim$ 20 Myr, ][]{Barcos-Munoz_2017}. 
We note that these regions with short depletion times are generally associated with the existence of young massive star clusters (YMCs; Fig. \ref{fig:33GHz}, magenta circles). We selected these young star clusters from \citet{Linden_2017} with age smaller than 10 Myr and mass greater than $10^6$ $\si{M_{\odot}}$. For example the south continuum source is associated with a YMC with age of 3.3 Myr and mass of $\sim 10^7\ \si{M_{\odot}}$. This is consistent with X-ray observations done by \citet{Smith_2014} who suggest the presence of a compact star forming region there. On the contrary, the center of NGC 5257 and the south arm of NGC 5258 do not have associated YMCs. We will discuss how this phenomenon and various other factors drive the difference in depletion time in Section 5. We also note there is an offset between the location of the star cluster and that of the continuum peak, such as in the south continuum source of NGC 5257. The offset for that particular source is 1.1 arcsec. We suspect this offset may be due to the inaccurate position registration of HST image.

In later sections, we will use the \cotwo\ map to trace the gas surface density assuming a typical \cotwo  /1-0 ratio. We generally assume the typical ratio of 0.8, which is the global flux ratio between the \cotwo\ and \coone\ map.

\section{Radiative Transfer Analysis}

\subsection{RADEX Modeling}

\begin{figure*}
	\centering
	\subfloat{\includegraphics[width=0.4\linewidth]{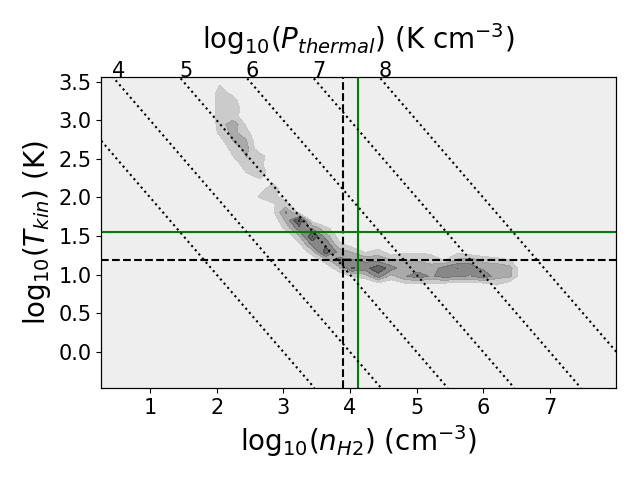}}%
	\subfloat{\includegraphics[width=0.4\linewidth]{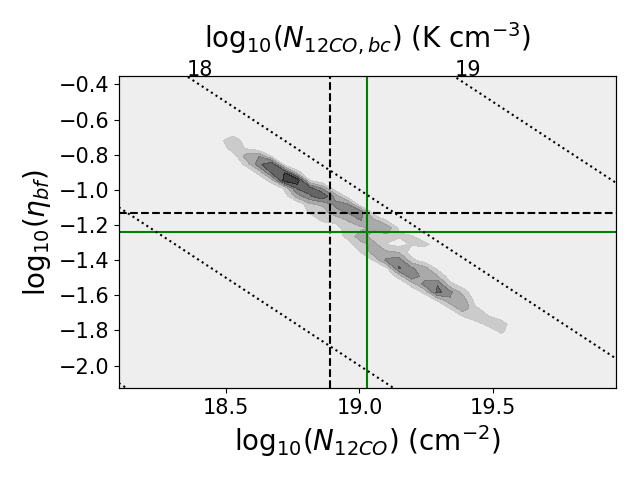}} \\
	\hspace*{-0.7cm}  
	\raisebox{10pt}{\subfloat{\includegraphics[width=0.4\linewidth]{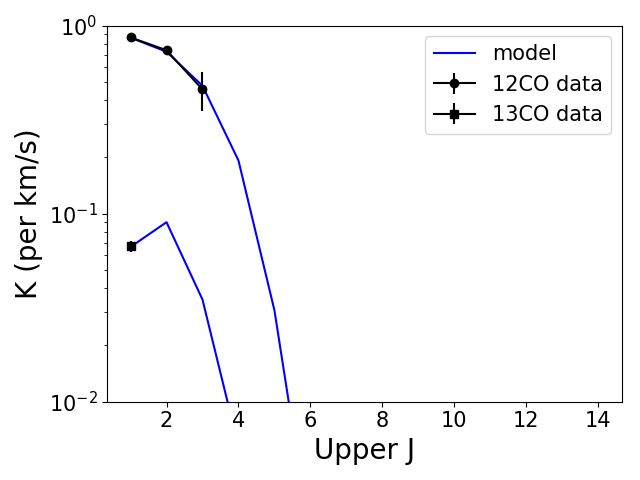}}}%
	\subfloat{\includegraphics[width=0.35\linewidth]{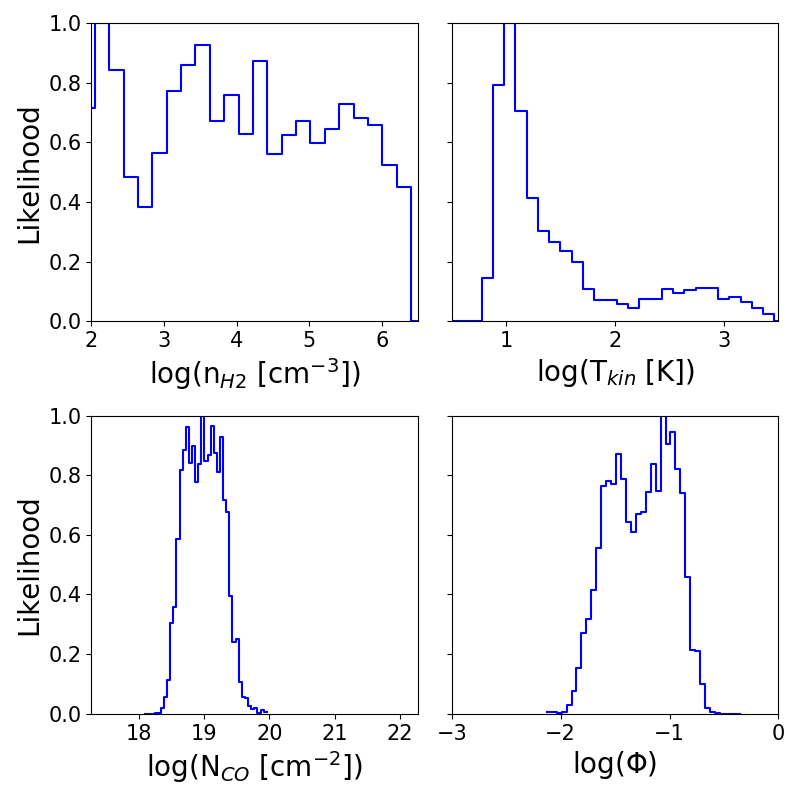}}
	\caption[RADEX modeling results of the center of NGC 5257]{RADEX modeling results for the center of NGC 5257 assuming \abun=50. (a) The temperature vs volume density probability distribution contours. Diagonal dot-dashed lines indicate constant thermal pressure; values of log($P$) are given along the top axis. The green and dashed crosses indicate the 1D mean value and 4D best fit values respectively.  (b) The column density vs beam filling factor probability distribution contour. Diagonal dot-dashed lines indicate constant beam averaged column density; values of log$(N_{\text{12CO, bc}})$ are given along the top axis. (c) The spectral line energy distribution (SLED) of the data and modeled result. (d) The 1D probability distribution function of temperature, volume density, column density and beam filling factor.  }
	\label{fig:NGC5257_radex}
\end{figure*}

\begin{figure*}
	\centering
	\subfloat{\includegraphics[width=0.4\linewidth]{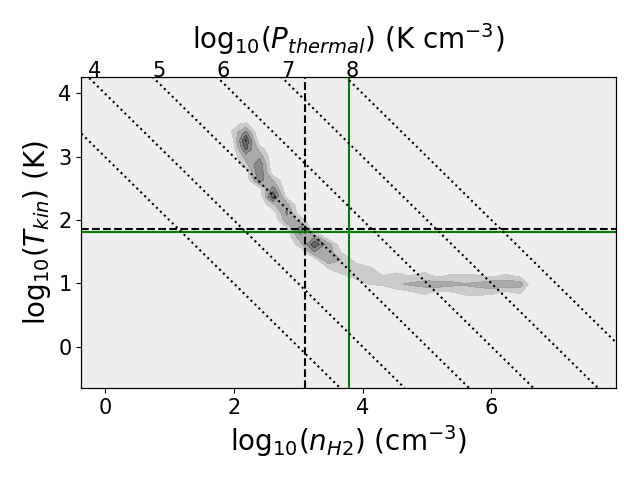}}%
	\subfloat{\includegraphics[width=0.4\linewidth]{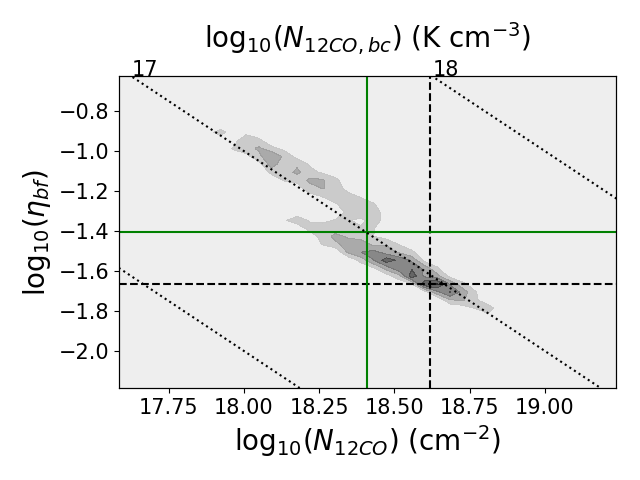}} \\
	\hspace*{-0.7cm}  
	\raisebox{10pt}{\subfloat{\includegraphics[width=0.4\linewidth]{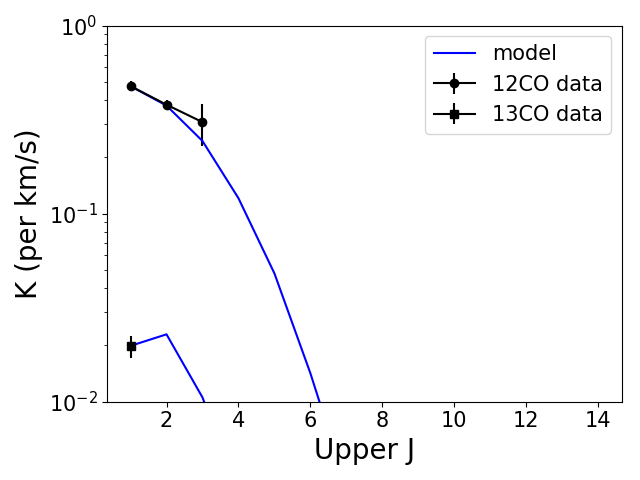}}}%
	\subfloat{\includegraphics[width=0.35\linewidth]{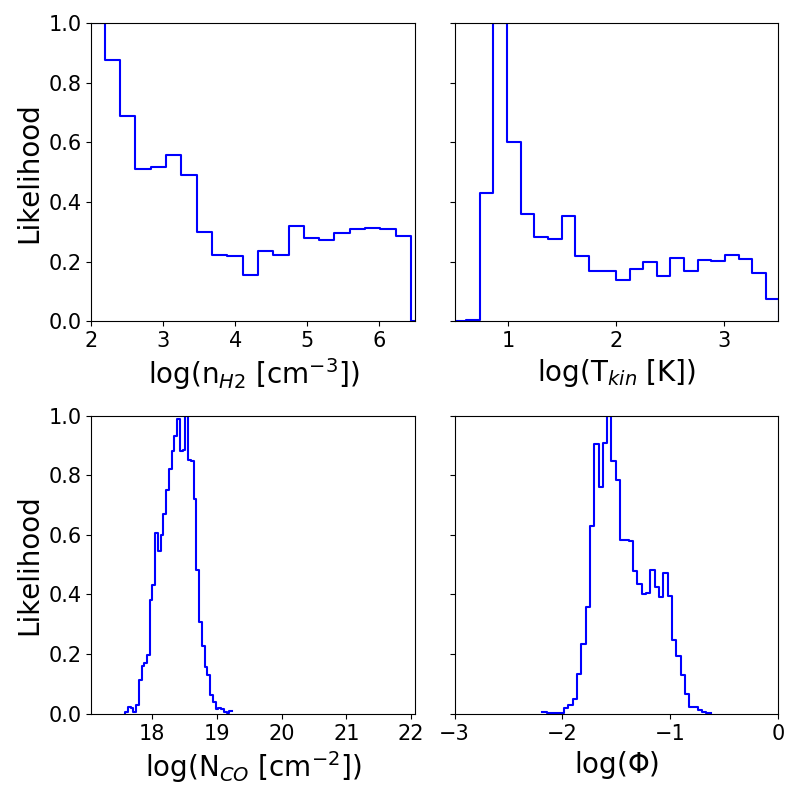}}
	\caption[RADEX modeling results of the south arm of NGC 5257]{RADEX modeling results for the south-west \cothree\ concentration region of NGC 5257 assuming \abun =50. See Figure \ref{fig:NGC5257_radex} for details. }
	\label{fig:NGC5257_radex_co32arm}
\end{figure*}

\begin{figure*}
	\centering
	\subfloat{\includegraphics[width=0.4\linewidth]{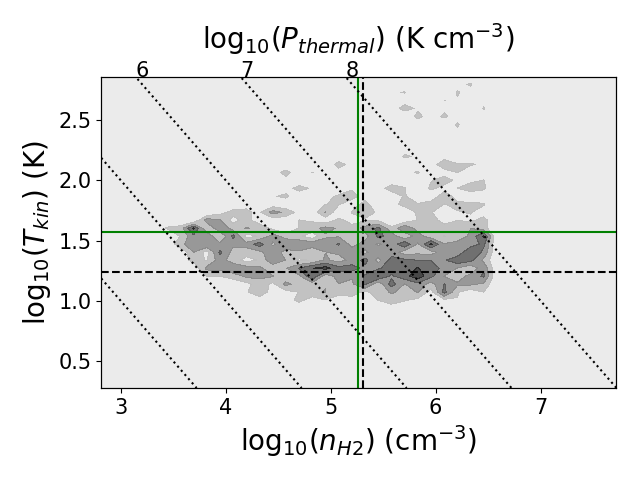}}%
	\subfloat{\includegraphics[width=0.4\linewidth]{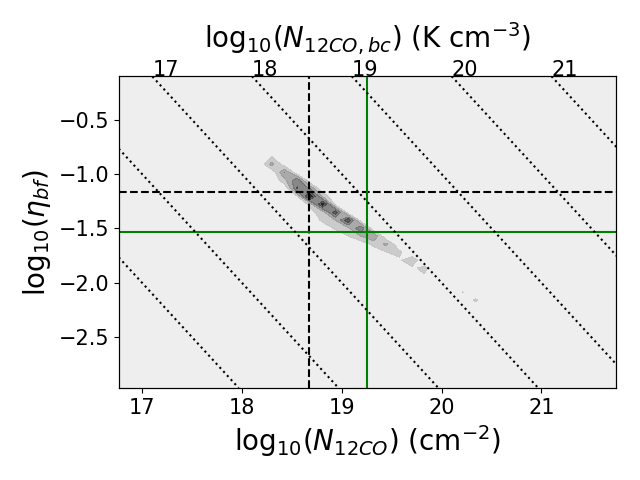}} \\
	\hspace*{-0.7cm}  
	\raisebox{10pt}{\subfloat{\includegraphics[width=0.4\linewidth]{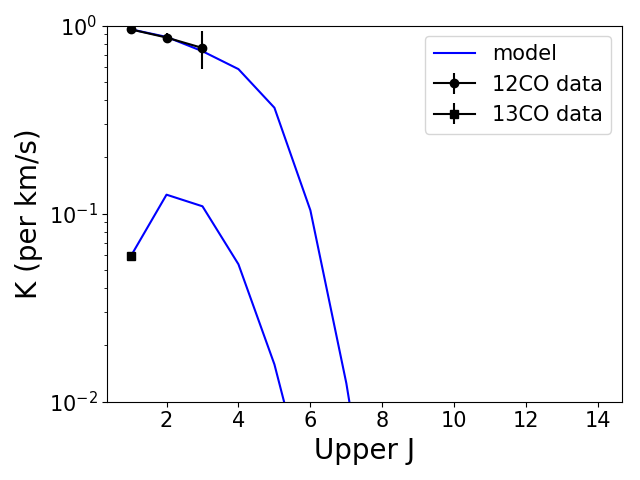}}}%
	\subfloat{\includegraphics[width=0.35\linewidth]{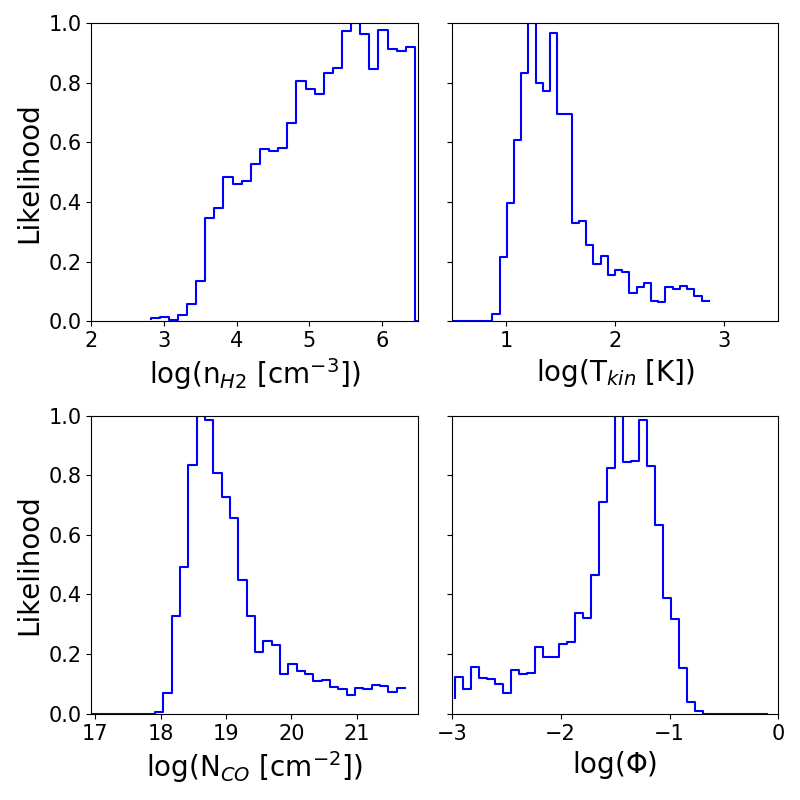}}
	\caption[RADEX modeling results for the south arm of NGC 5258]{RADEX modeling results of the south arm of NGC 5258 assuming \abun =50. See Figure \ref{fig:NGC5257_radex} for details. }
	\label{fig:NGC5258_radex}
\end{figure*}

To constrain the physical properties of the molecular gas, we use the radiative transfer code RADEX \citep{vandertak_2007}. This code calculates line intensity based on input temperature, number density of \htwo\ and column density of the modeled molecules divided by the linewidth.
We use a grid of models across the parameter space to fit the observed line intensities. In addition to gridding, we use a Baysian likelihood code \citep{Kamenetzky_2016} to create probability distributions for the various parameters above. This code also introduces an additional parameter called the beam filling factor, which is how large a fraction of the area CO emission actually covers within a single beam. This factor is somewhat degenerate with the actual CO column density, but these two parameters have different effects on the CO line ratios.
This code gives two types of solutions, 1D Max  and 4D max. The 1D max gives the solution of the parameters with maximal likelihood in one dimensional parameter space. The 4D max gives the solution with maximal likelihood for the combination of all 4 parameters listed above.

We use all 3 CO lines observed with ALMA and the \cothree\ line observed with SMA to model regions with \cothree\ detections (Fig. \ref{fig:SMA}). Therefore we smooth all the line images to the largest beam size of 3.8\arcsec\ $\times$ 2.99\arcsec\ and measure the average intensity. Possibly due to the missing flux problem of the SMA data \citep{Wilson_2008}, the center of NGC 5257 has an extremely low \cothree\ flux compared to the fluxes of the other lines. Therefore, we use the peak intensity instead of the average intensity for the center of NGC 5257. To constrain all the parameters, we would still need an additional \tco\ line.  
Therefore, we chose to fix the \abun\ abundance ratio to some certain values and run the model. For the Milky Way, this ratio varies from $\sim$ 30 in the center to $\sim$ 100 in the outermost disk \citep{Milam_2005}. We thus set the fiducial value for \abun\ to be 50, and the lower and upper extreme values to be 30 and 100. Because we only model the lower J CO lines, we use a one-component model. This model also requires the linewidth for each region. For this quantity we measured the median full width at half maximum (FWHM) of the lines in the selected regions, which are 135, 116 and 87 km s$^{-1}$ for the center and south-west region of NGC 5257 and the south arm of NGC 5258 respectively. Before the modeling, we set some prior assumptions and conditions, which are listed below. 
\begin{itemize}
	\item We assume the [\co]/[\htwo] abundance ratio to be $10^{-4}$ \citep{Cormier_2018} with a helium correction to be 1.4. 
	\item We set the prior limit to the optical depth within the range of [0, 100]. The lower limit is set due to the fact that CO should not be a maser. The upper limit is recommended by the RADEX documentation.
\end{itemize}

Some outputs from the modeling for \abun\ = 50 are shown in Fig. \ref{fig:NGC5257_radex}, \ref{fig:NGC5257_radex_co32arm} and \ref{fig:NGC5258_radex} for illustration. From these figures, we can see the modeling generally reproduces the measured CO SLEDs. We can also see from the 2D contour plots that there are not significant differences between the 1D solution and the 4D solution, with a maximal difference of 0.5 dex. For more details about these figures, see the caption in Fig. \ref{fig:NGC5257_radex}.
The 1D solutions for different regions with different \abun\ assumptions are shown in Table \ref{tab:radex_result}. Most results vary with the different assumed \abun\ ratios. The density varies by 2 orders of magnitude among different \abun\ ratios and cannot be constrained very well. Temperature, column density and beam filling factor vary less. 

We can see for both the center of NGC 5257 and the south arm of NGC 5258, where the gas is mostly concentrated, the temperature and the density are similar. However, for the south-west \cothree\ luminous region in NGC 5257, the temperature is significantly higher and the density is significantly lower. This is what we expect as this region does not show an emission peak in the \coone\ and \cotwo\ moment 0 maps. The temperature is generally within the range of 10--100 K, which is reasonable for molecular gas.

	\begin{table*} 
		\centering
		\caption{RADEX Modeling Results}
        \begin{threeparttable}
		\renewcommand{\arraystretch}{1.5}
		\begin{tabular}{lcccccccc} 
			\hline
			Galaxy & Region & Assumed & $T_{\text{kin}}$ $^b$ & log$(n_{\text{H2}}) $ $^c$ & log$(N_{\text{12CO}})$ $^d$ & $\eta_{\text{bf}}$ $^e$ & log($P$) $^f$ & log$(N_{\text{12CO,bc}})$ $^g$ \\
			& & \abun\ $^{a}$ & (K) & (cm$^{-3}$) & (cm$^{-2}$) & & (K cm$^{-3}$) & (cm$^{-2}$) \\
			\hline
			NGC 5257 & Center & 100 & $66\ _{-49}^{+183}$ & $3.2
			\pm 1.3$ & $19.54 \pm 0.31 $ & $0.04 _{-0.02}^{+0.04}$ & $5.1 \pm 0.9$ & $18.12 \pm 0.10$ \\
			NGC 5257 & Center & 50 & $35\ _{-28}^{+131}$ & $4.1 \pm 1.3$ & $19.03 \pm 0.31 $ & $0.06\ _{-0.03}^{+0.06}$ & $5.7\pm 0.8 $ & $17.79 \pm 0.10$ \\
			NGC 5257 & Center & 30 & $19\ _{-12}^{+36}$ & $4.5 \pm 1.1$ & $18.61 \pm 0.24$ & $0.09\ _{-0.04}^{+0.07}$ & $5.8 \pm 0.8 $ & $17.55 \pm 0.07$ \\
			NGC 5257 & South West & 100 & $94\ _{-76}^{+476}$ & $3.6 \pm 1.4$ & $18.98 \pm 0.36$ & $0.02\ _{-0.01}^{+0.02}$ & $5.5 \pm 0.9 $ & $17.33 \pm 0.14$ \\
			NGC 5257 & South West & 50 & $64\ _{-54}^{+359}$ & $3.8 \pm 1.4$ & $18.40 \pm 0.28 $ & $0.04\ _{-0.01}^{+0.03}$ & $5.6\pm 0.8 $ & $17.00 \pm 0.11$ \\
			NGC 5257 & South West & 30 & $314\ _{-250}^{+1218}$ & $2.7
			\pm 0.7$ & $17.71 \pm 0.30 $ & $0.10 _{-0.04}^{+0.08}$ & $5.2 \pm 0.4$ & $16.74 \pm 0.06$
			\\
			NGC 5258 & South Arm & 100 & $56\ _{-38}^{+115}$ & $5.0 \pm 1.0$ & $19.77\pm 0.84$ & $0.02\ _{-0.01}^{+0.04}$ & $6.8\pm 1.0$ & $18.08 \pm 0.38 $ \\	
			NGC 5258 & South Arm & 50 & $37\ _{-23}^{+63}$ & $5.3 \pm 0.8$ & $19.25 \pm 0.83$ & $0.03\ _{-0.02}^{+0.06}$ & $6.8\pm 1.0$ & $17.72 \pm 0.36$ \\
			NGC 5258 & South Arm & 30 & $17\ _{-8}^{+14}$ & $5.1 \pm 0.9$ & $18.33 \pm 0.42$ & $0.08\ _{-0.04}^{+0.08}$ & $6.4 \pm 0.9$ & $17.26 \pm 0.16$ \\
			\hline
		\end{tabular}
		\end{threeparttable}
		\begin{tablenotes}
			\item \textbf{Columns}: $^{(a)}$\abun\ abundance ratio. $^{(b)}$The kinetic temperature of the gas. $^{(c)}$The volume density of hydrogen. $^{(d)}$The \co\ column density. $^{(e)}$The \co\ beam filling factor. $^{(f)}$The thermal pressure of gas, which is the product of the kinetic temperature and volume density. $^{(g)}$The \co\ beam averaged column density, which is the product of the \co\ column density and beam filling factor. 
		\end{tablenotes}
		\label{tab:radex_result}
	\end{table*}

For each assumed \abun\ ratio value, the beam averaged column density, which is the multiplication of column density and beam filling factor, is more accurately modeled than the two individual parameters. Therefore, we will use the beam averaged column density to calculate the CO-to-\htwo\ conversion factor in different regions.

\subsection{CO-to-\htwo\ Conversion Factor}

From the RADEX modeling, we get the 1D mean result for the \co\ beam averaged column density. We can calculate the H$_2$ surface density based on the \co\ column density of the gas and an assumed [\co]/[\htwo] abundance ratio with
\begin{equation}
N_{H_2}=\frac{N_{\text{12CO}}} {[^{12}\mathrm{CO}]/[\mathrm{H}_2]}
\end{equation}
where $N_{H_2}$ is the column density of H$_2$, $N_{\text{12CO}}$ is the column density of $^{12}$CO and $[^{12}\mathrm{CO}]/[\mathrm{H}_2]$ is the CO to \htwo\ abundance ratio. In this case we assume $[^{12}\mathrm{CO}]/[\mathrm{H}_2]=1 \times 10^{-4}$ \citep{Cormier_2018}. We can compare these results directly with the \coone\ intensity to calculate the conversion factor. 

We calculate \alphaco\ for these three regions separately. From the calculation, we can see the conversion factors of the three regions are closer to the typical ULIRG conversion factor. However, the derived conversion factor is highly dependent on the assumed \abun\ and [\co]/[\htwo] abundance ratios. To constrain the \abun\ ratio, we need at least one more \tco\ line.

\begin{table}
	\centering
	\caption{RADEX modeled \alphaco\ results}
	\renewcommand{\arraystretch}{1.5}
	\begin{threeparttable}
	\small\addtolength{\tabcolsep}{-2pt}
	\begin{tabular}{lcccc}
		\hline
		Galaxy   & Region    & \multicolumn{3}{c}{\alphaco\  ($\si{M_{\odot}\ pc^{-2}\ (K\ km\ s^{-1}})^{-1}$)}                       \\
		&           & \abun & ... & ... \\ 
		& & =100 & =50 & =30 \\ \hline
		NGC 5257 & Center    & $1.9_{-0.4}^{+0.5}$           & $0.9_{-0.2}^{+0.2}$              & $0.5_{-0.1}^{+0.1}$               \\
		& South West  & $0.9_{-0.3}^{+0.4}$ & $0.4_{-0.1}^{+0.1}$ &   $0.23_{-0.03}^{+0.04}$ \\
		NGC 5258 & South Arm & $3.4_{-2.0}^{+4.7}$                & $1.5_{-0.8}^{+1.9}$               & $0.5_{-0.2}^{+0.3}$               \\ \hline
	\end{tabular}
	\end{threeparttable}
	\label{tab:conversion}
\end{table}

\subsection{Comparison with LTE Analysis}

\begin{table}
	\centering
	\caption{LTE calculated mean \alphaco\ results}
	\begin{threeparttable}
	\renewcommand{\arraystretch}{1.2}
	\begin{tabular}{lcc}
		\hline
		Galaxy                    & \multicolumn{2}{c}{Conversion Factor ($\si{M_{\odot}\ pc^{-2}\ (K\ km\ s^{-1}})^{-1}$)} \\
		 & $\eta_{\text{bf}}$=1.0 & $\eta_{\text{bf}}$=0.1 \\  \hline
		NGC 5257 & 0.50                            & 0.94                           \\
		NGC 5258                  & 0.59                                             & 1.27                                            \\ \hline
	\end{tabular}
	\end{threeparttable}
	\begin{tablenotes}
		\item a. We assume \abun=50
		\item b. The value is the mass weighted mean value
	\end{tablenotes}
	\label{tab: LTE_alpha}
\end{table}

We use a local thermal equilibrium (LTE) analysis for the \co/\tcoone\ ratio to try to understand what causes the change in the ratio. When both \co\ and \tco\ lines are optically thin, the ratio of the brightness temperatures will tell us about the abundance ratio. In the general case, \co\ is optically thick while \tco\ is optically thin. In that case \citep{Cormier_2018}, 
\begin{equation}
R=\frac{T_{\mathrm{ex, 12}}(1-e^{-\tau_{\mathrm{12CO}}})}{T_{\mathrm{ex, 13}}(1-e^{-\tau_{\mathrm{13CO}}})}
\end{equation}
where $(1-e^{-\tau_{\mathrm{12CO}}}) \rightarrow 1$ and $(1-e^{-\tau_{\mathrm{13CO}}}) \rightarrow \tau_{\mathrm{13CO}}$. The ratio can then be simplified as 
\begin{equation}
R= \frac{1}{\tau_{\mathrm{13CO}}}
\end{equation}
We calculate the column density based on the excitation temperature and optical depth via
\begin{equation}
N_{13}=\frac{3.0\times 10^{14}}{1-\exp(-5.29/T_{ex,13})}\times \frac{\tau_{\mathrm{13CO}}}{1-\exp(-\tau_{\mathrm{13CO}})}\times I_{13CO} [\si{cm^{-2}}]
\end{equation}
where $N_{13}$ is the \tco\ column density in $\si{cm^{-2}}$ and $I_{13CO}$ is the \tco\ intensity in the units of $\si{K\ km\ s^{-1}}$. Under the LTE assumption, the excitation temperature of \co\ and \tco\ should be equal. We can then calculate the \co\ excitation temperature $T_{\mathrm{ex, 12}}$ as in  \citet{Heiderman_2010} assuming $\tau_{\mathrm{12CO}} \rightarrow \infty$. 
\begin{equation}
T_{\mathrm{ex, 13}}=T_{\mathrm{ex, 12}}=\frac{5.5}{\ln\left(1+\frac{5.5}{T_{\mathrm{peak, 12CO}}/\eta_{\text{bf}}+0.82}\right)}
\end{equation}
where $\eta_{\text{bf}}$ is the beam filling factor of \co\ and $T_{\mathrm{peak, 12CO}}$ is the \coone\ peak main beam brightness temperature. Compared to the original equation, we add the beam filling factor term as we do not have giant molecular cloud scale resolution. 


We use these equations to map the conversion factor across both galaxies. We assume the abundance ratio of \abun\ to be 50. We assume $\eta_{\text{bf}}$ to be 0.1 and 1.0 respectively to see how this term affects our results. We also assume that [\co]/[\htwo]=$10^{-4}$, which is the same as our input into the RADEX model. We then binned the pixels to be 1.5\arcsec $\times$ 1.5\arcsec. From the map of LTE \alphaco\ for each pixel, we calculated the mean \alphaco\ weighted by calculated $N_{\text{12CO}}$. The results are shown in Table \ref{tab: LTE_alpha}. We can see within the range of reasonable $\eta_{\text{bf}}$, the mean value of \alphaco\ for both galaxies are close to the ULIRG value, which is consistent with our RADEX modeling results. 

\subsection{Factors impacting the $\mathbf{\alpha_{\text{CO}}}$ value}

By using RADEX modeling, we derived the \alphaco\ for three regions of Arp 240 with \cothree\ detection. We find that all of these regions seem to have \alphaco\ close to the commonly used ULIRG value (1.1 \alphacou\ including helium). This is also supported by LTE analysis across the entire galaxies. There are several prior assumptions that need to be considered while interpreting the results. The first assumption is the \abun\ abundance ratio. This value is poorly constrained in studies of U/LIRGs and varies among different merging stages. Post mergers, such as NGC 2623, have \abun\ ratios exceeding 200 while early mergers, such as Arp 55 have \abun\ ratios of 15 $\sim$ 30 in both nuclei \citep{Sliwa_2017}. If we assume Arp 240 has low \abun\ like Arp 55, we would expect \alphaco\ values to be smaller than 0.5 \alphacou, which is even lower than the commonly used ULIRG value. The other factor that comes in is the assumed [\co]/[\htwo] abundance ratio. 
In this case, we assume the [\co]/[\htwo] abundance ratio to be 1$\times 10^{-4}$. Clouds in the Milky Way generally have [\co]/[\htwo] of about $ 10^{-4}$ \citep{vanDishoeck_1992} with values ranging from $5 \times 10^{-5}$ to $2.7 \times 10^{-4}$ \citep[references in][]{Zhu_2003}. \citet{Zhu_2003} derived the typical [\co]/[\htwo] value for the Antennae of between $0.5 \sim 1.0 \times 10^{-4}$. However, \citet{Sliwa_2017} assumed [\co]/[\htwo] $\sim 3 \times 10^{-4}$ for U/LIRGs. They argue that this is typical value for warm star forming clouds, which should commonly exist in U/LIRGs. If this is the case for Arp 240, then our estimated conversion factors should be even smaller. All of this evidence suggests that this merger should have an \alphaco\ smaller or comparable to the  typical ULIRG conversion factor.

The result is consistent with simulations of mergers, such as in \citet{Renaud_2019}. According to their simulations, the \alphaco\ of an Antennae-like galaxy drops to its lowest value 30 -- 80 Myrs after first pericentric passage. Arp 240 is a similar major merger and it has just been
about 220 Myr after its first pericentric passage. So Arp 240 might still have relatively low \alphaco\ value due to on-going star forming activities. 

The cause for such a low \alphaco\ is not clear in our case. Both gas concentration regions, the center of NGC 5257 and the south arm of NGC 5258, have a relatively low temperature (Table \ref{tab:radex_result}), which seems to exclude the temperature influence. Since we don't have GMC resolution ($\sim$ 100 pc) of these regions, we can't explicitly determine if velocity dispersion plays an important role. On the other hand, the \cothree\ emission in the south-west arm seems to trace the hot diffuse gas, as no emission peak found in \coone\ and \cotwo\ maps. This is probably why it has the lowest \alphaco\ compared to the other regions.

\section{Gas Properties and SFR}

\begin{figure*}
	\centering
	\subfloat{\includegraphics[width=0.5\linewidth]{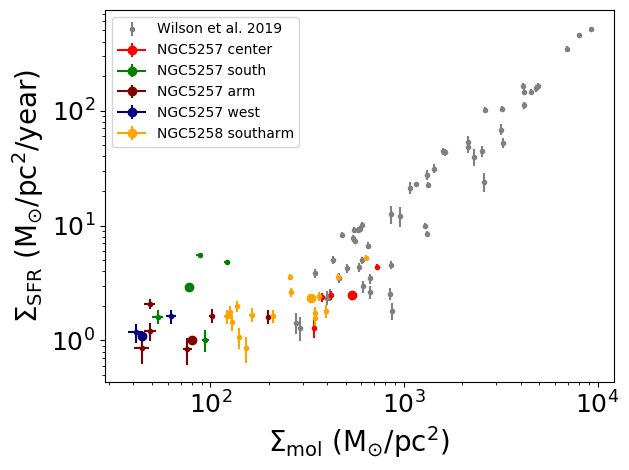}}%
	\subfloat{\includegraphics[width=0.5\linewidth]{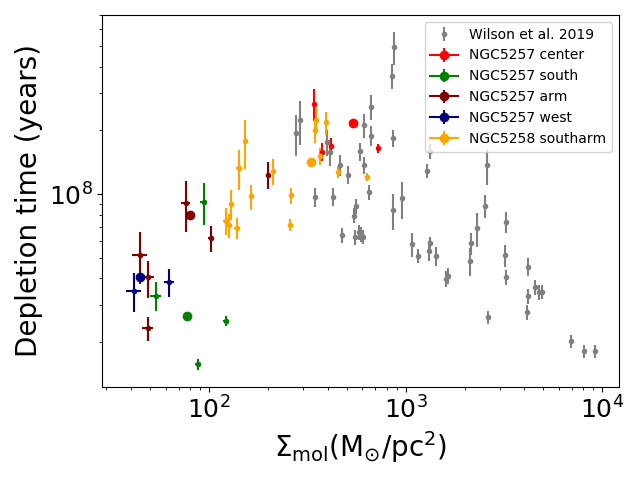}} 
	\caption[$\Sigma_{\text{SFR}}$, $t_{\text{dep}}$ and $\epsilon_{\text{ff}}$ versus $\Sigma_{\text{mol}}$ for NGC 5257 and NGC 5258]{(Left) $\Sigma_{\text{SFR}}$ versus $\Sigma_{\text{mol}}$ . (Right) $t_{\text{dep}}$ versus $\Sigma_{\text{mol}}$. We assume the ULIRG conversion factor for all the galaxies. We do not correct for inclination in calculating the surface density. The big dots represent the average of pixels in the different polygon regions shown in Fig. \ref{fig:33GHz}. }
	\label{fig:KS}
\end{figure*}

We use the 33 GHz continuum image to trace the SFR. We divide both galaxies into different regions based on the 33 GHz continuum image, as shown in Figure \ref{fig:33GHz}. We can then calculate the depletion time with
\begin{equation}
t_{\text{dep}}=M_{\text{mol}}/\mathrm{SFR} = \Sigma_{\text{mol}}/\Sigma_{\text{SFR}}
\end{equation}
The molecular mass is calculated based on the \cotwo\ intensity. We assume the conversion factor to be a ULIRG conversion factor. We also assume the \cotwo/1-0 ratio is 0.8, which is the flux ratio measurement from section 3.2. The depletion times are shown in Table \ref{tab:tdep}. We can see different regions have very different depletion times. We plotted the SFR and depletion time as a function of molecular gas surface density for each region, as shown in Fig. \ref{fig:KS}. To show if there is a clear trend in any of these plots, we also added points for the individual pixels, which are shown as smaller points with different colors indicating which region they come from.  To measure the SFR and $\Sigma_{\text{mol}}$ for each individual pixel, we first smooth both the \cotwo\ and 33 GHz images to a beam size of 1.1\arcsec\ $\times$ 0.8\arcsec. We then binned the pixels to 1.0\arcsec\ $\times$ 1.0\arcsec to get independent measurements. For pixel selection, we applied a S/N=4 cut to the 33 GHz image. We also add in resolved points from other ULIRGs \citep{Wilson_2019} for comparison. 

We can see that the SFR and $t_{\text{dep}}$ extend smoothly from the results of U/LIRGs. However, we can see the left plot in Fig. \ref{fig:KS} shows a flattening trend as $\Sigma_{\text{mol}}$ decreases. This is probably caused by selection bias as the radio continuum is not sensitive to faint star forming regions. This cut off is also reflected in the right subplot of the same figure. As $\Sigma_{\text{mol}}$ decreases, we constantly have points with SFR just above the detection limit. The relatively constant SFR divided by decreasing $\Sigma_{\text{mol}}$ will give us the decreasing trend of $t_{\text{dep}}$ as $\Sigma_{\text{mol}}$ decreases. However, the aperture measurements follow a similar trend as the individual pixels. In the following subsections, we will try to interpret these plots from different perspectives.

\subsection{SFE per free-fall time} 

\begin{figure*}
	\centering
	\subfloat{\includegraphics[width=0.5\linewidth]{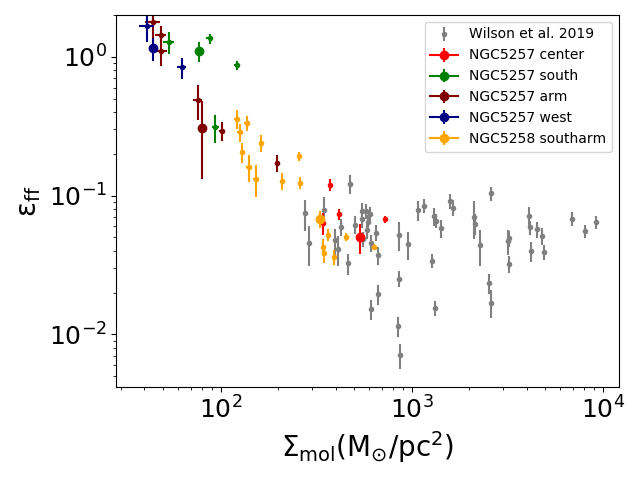}}%
	\subfloat{\includegraphics[width=0.5\linewidth]{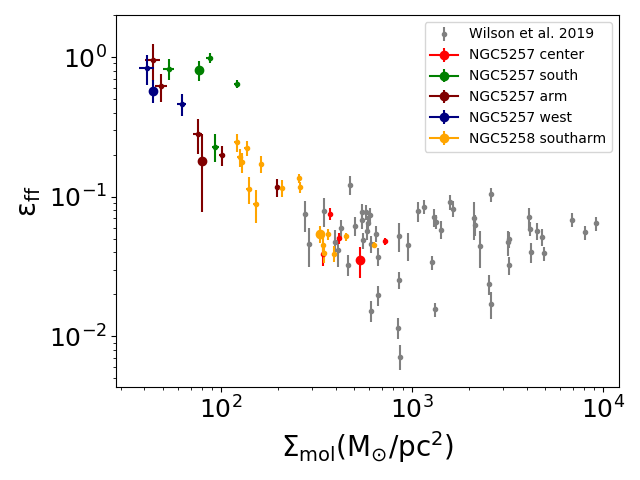}} \\  
	\caption{SFE per free-fall time with fixed (left) and varying gas fraction (right) of gas component as a function of $\Sigma_{\text{mol}}$. The symbols are the same as in Fig. \ref{fig:KS}. }
	\label{fig:epsff_stellar}
\end{figure*}

One of the theoretical models to account for different depletion times is the turbulence model of \citet{Krumholz_Mckee_2005}. According to this model, turbulence will determine the density PDF of the clouds. The most important parameter is the SFE per free-fall time, which is the ratio between the free-fall time and the depletion time. The general equation for free-fall time is 
\begin{equation}
t_{\text{ff}}=\sqrt{\frac{3\pi}{32G\rho_{\mathrm{mol, mid}}}}
\end{equation}
where $\rho_{\mathrm{mol, mid}}$ is the volume density of the molecular gas in the middle of the disk. If we assume the galaxy is filled with gas, then the volume density can be calculated as
\begin{equation}
\rho_{\mathrm{mol, mid}}=\frac{\Sigma_{\text{mol}}}{2 H_{\mathrm{mol}}}
\end{equation}
where $H_{\mathrm{mol}}$ is the scale height of the disk and $\Sigma_{\text{mol}}$ is the surface density of the molecular gas measured from the \cotwo\ cube (see Section 3.2). Assuming the gas disk is in equilibrium and that vertical gravity is dominated by gas self-gravity, then we can calculate the scale height as \citep{Wilson_2019}
\begin{equation}
H_{\mathrm{mol}}= 0.5\frac{\sigma_v^2}{\pi G \Sigma_{\text{mol}}}
\end{equation}
where ${\sigma_v}$ is the velocity dispersion of the molecular gas.
Combining all the equations above, we can write the free-fall time as 
\begin{equation}
t_{\text{ff}}=\frac{\sqrt{3}}{4G} \frac{\sigma_v}{\Sigma_{\text{mol}}}
\end{equation}
Therefore the SFE per free-fall time can be calculated as \citep{Wilson_2019}
\begin{equation}
\label{equ: epsff1}
\epsilon_{\text{ff}}=\frac{t_{\text{ff}}}{t_{\text{dep}}}=\frac{\sqrt{3}}{4G}\frac{\sigma_v \Sigma_{\text{SFR}}}{\Sigma^2_{\mathrm{mol}}}
\end{equation}
Assuming a constant SFE per free-fall time, the depletion time will decrease as the surface density of the gas increases. To calculate $\epsilon_{\text{ff}}$, we need maps of $\Sigma_{\text{mol}}$, $\Sigma_{\text{SFR}}$ and $\sigma_v$. We plot $\epsilon_{\text{ff}}$ versus $\Sigma_{\text{mol}}$ in Fig. \ref{fig:epsff_stellar} (left plot).
As we can see, there is a clear decreasing trend for $\epsilon_{\text{ff}}$ versus $\Sigma_{\text{mol}}$ for Arp 240, while $\epsilon_{\text{ff}}$ stays relatively constant for the other U/LIRGs. Furthermore, in the low surface density regions, $\epsilon_{\text{ff}}$ can be above 1.0. For normal clouds, $\epsilon_{\text{ff}}$ can be as high as several 10\% \citep{Lee_2016}. Therefore, our method probably overestimates the true efficiency in these galaxies. 

In the above calculation, we assume gas collapse is dominated by its self-gravity. However, the stellar component might play an important role in the collapse of the gas disk. As we have measurement of the stellar surface density from \textit{Spitzer} near infrared data, we can try to estimate the gas fraction within the scale height of the gas disk. As we do not have direct measurement of the stellar velocity dispersion, we use the empirical equation from \citet{Leroy_2008} to estimate the scale height for the stars, which is
\begin{equation}
H_{\star}=\frac{l_{\star}}{7.3}
\end{equation}
where $l_{\star}$ is the scale length of the stellar disk, which is 4.2 kpc and 5.8 kpc for NGC 5257 and NGC 5258 respectively \citep{Fuentes-Carrera_2019}. Therefore, we can get the stellar midplane density, which is 
\begin{equation}
\rho_{\star, \mathrm{mid}}=\frac{\Sigma_{\star}}{2 H_{\star}}
\end{equation}
The stellar component also affects the gas scale height and thus the gas midplane density. We correct for this effect in Appendix \ref{epsff_star} and get the modified midplane gas density $\rho_{\mathrm{mol, mid}}^*$. Therefore, we can calculate the midplane gas fraction as 
\begin{equation}
f_{\mathrm{mid}}=\frac{\rho_{\mathrm{mol, mid}}^*}{\rho_{\star, \mathrm{mid}}+\rho_{\mathrm{mol, mid}}^*}
\end{equation} 
and the SFE per free-fall time becomes
\begin{equation}
\epsilon_{\mathrm{ff, mod}}=\epsilon_{\text{ff}}\sqrt{\frac{f_{\mathrm{mid}}}{0.5}}
\end{equation}
The detailed derivation is given in Appendix \ref{epsff_star}. The comparison of the two efficiencies per free-fall time is shown in Figure \ref{fig:epsff_stellar}. We can see that the efficiency per free-fall time including the stellar component is slightly smaller. However, at low gas surface densities, there are still a lot of points with SFE per free-fall time $\sim$ 100\%. 

Therefore, we still need some explanation for the extreme value of $\epsilon_{\text{ff}}$. We note that these extreme points are generally from the three off-center regions in NGC 5257. As identified in \citet{Linden_2017}, these regions are generally associated with young massive star clusters (YMCs). Among them, the south of NGC 5257 has the shortest depletion time, as shown in Fig. \ref{fig:KS}. This region is particularly associated with a young ($\sim$ 3.3 Myr) and massive ($\sim$ 10$^7\ \si{M_{\odot}}$) star cluster. We will further discuss how this connects to the extreme $\epsilon_{\text{ff}}$ value in Section 5.3.

\subsection{Toomre Stability}

\begin{figure*}
	\centering
	\subfloat{\includegraphics[width=0.5\linewidth]{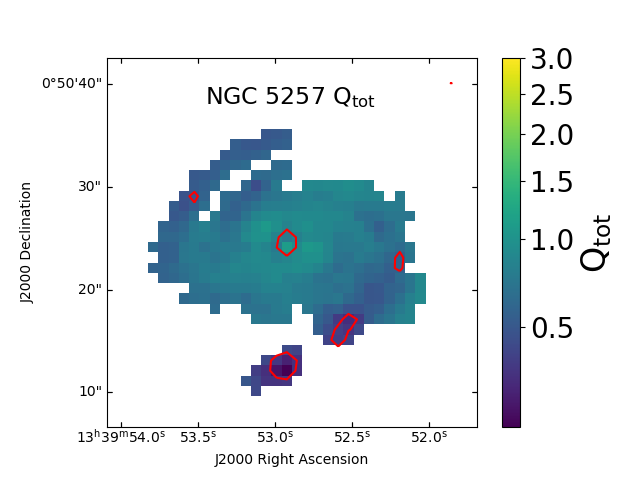}}%
	\subfloat{\includegraphics[width=0.5\linewidth]{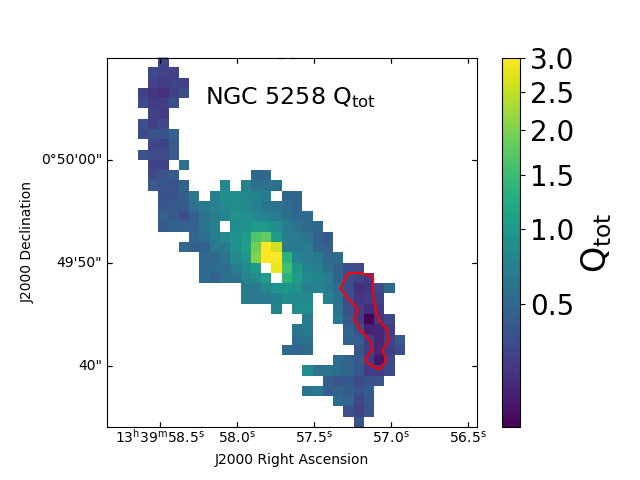}} 
	\caption[Toomre factor map (left) and scatter plot (right) of NGC 5257 and NGC 5258]{The Toomre Q factor map of NGC 5257 (left) and NGC 5258 (right). The red contour in the map is the 33 GHz continuum at a level of 2.0 $\times 10^{-5}\ \si{Jy\ beam^{-1}}$. }
	\label{fig:Toomre_map}
\end{figure*}

As shown in Fig. \ref{fig:vel}, both galaxies are still fairly normal rotating disks. Therefore, we can apply a Toomre stability analysis \citep{Toomre1977} to see if the star forming regions satisfy the instability criterion and if there is a dependence of $t_{\text{dep}}$ on the value of the Toomre factor. Comparing to \citet{Leroy_2008}, we use the measured gas velocity dispersion instead of assuming a constant value. Therefore, we might get different conclusions from \citet{Leroy_2008}, who found there is no correlation between the Toomre factor and the depletion time. 

The Toomre factor for a pure gas disk is given by 
\begin{equation}
Q=\frac{\sigma_v \kappa}{\pi G \Sigma_{\text{mol}}}
\end{equation}
where $\Sigma_{\text{mol}}$ and $\sigma_v$ are the surface density and the velocity dispersion measured with the \cotwo\ line. $\kappa$ is the epicyclic frequency given by
\begin{equation}
\kappa=\sqrt{2}\frac{V}{R}(1+\frac{d\ln V}{d \ln R})
\end{equation}
where $V$ is the rotational speed and $R$ is the radius from the center. We use the parametric fitted rotation curve from  \citet{Fuentes-Carrera_2019}. For galaxies, the stellar gravitational potential will help to stabilize the disk. \citet{Wang_Silk_1994} introduce the composite Toomre factor which includes the effect of stars 
\begin{equation}
Q_{\mathrm{tot}}=Q(1+\frac{\Sigma_{\star}}{\Sigma_{\text{mol}}}\frac{\sigma_v}{\sigma_{\star}})
\end{equation}
where $\Sigma_{\star}$ and $\sigma_{\star}$ are the surface density and velocity dispersion of stars in the galaxy. The surface density of the stars is obtained with 3.6 \textmu m and 4.5 \textmu m \textit{Spitzer} image (see Section 2.3). We do not have data for $\sigma_{\star}$. We use the theoretical equation \citep{Leroy_2008} based on the assumption of disk hydrostatic equilibrium to calculate the stellar velocity dispersion,  
\begin{equation}
\begin{split}
&\sigma_{\star,z}=\sqrt{\frac{2\pi G l_{\star}}{7.3}}\Sigma_{\star}^{0.5} \\
&\sigma_{\star}=\sigma_{\star,R}=\sigma_{\star,z}/0.6
\end{split}
\end{equation}
where $\sigma_{\star,R}$ and $\sigma_{\star,z}$ are the stellar velocity dispersion along the vertical and radial direction and $l_{\star}$ is the scale length of the stellar disk, which is 4.2 kpc and 5.8 kpc for NGC 5257 and NGC 5258 respectively \citep{Fuentes-Carrera_2019}.

The $Q_{\mathrm{tot}}$ maps for the two galaxies are shown in Figure \ref{fig:Toomre_map}. Both galaxies have the majority of the disk with $Q_{\mathrm{tot}} < 1$, except for the very central region of NGC 5258. For the center of NGC 5257, the Toomre factor is close to the critical value of 1.0, which satisfies the instability criterion. Except for that region, all the star forming regions have a Toomre factor below 0.5, which is highly unstable to gravitational collapse. In the center of NGC 5258, the Q value is about 3, which is consistent with the fact that this is not a star forming region. We also calculated the $Q_{\mathrm{gas}}$ value for each pixel. $Q_{\mathrm{gas}}$ is larger than $Q_{\mathrm{tot}}$ by a factor of $2.5 \sim 10$ with most values greater than 1.0. This suggests the disk instability is mainly driven by stellar component. This is consistent with \citet{Romeo_2017}, who show the gas instability is mainly driven by stellar component in normal spiral galaxies. 

\begin{figure*}
	\centering
	\subfloat{\includegraphics[width=0.5\linewidth]{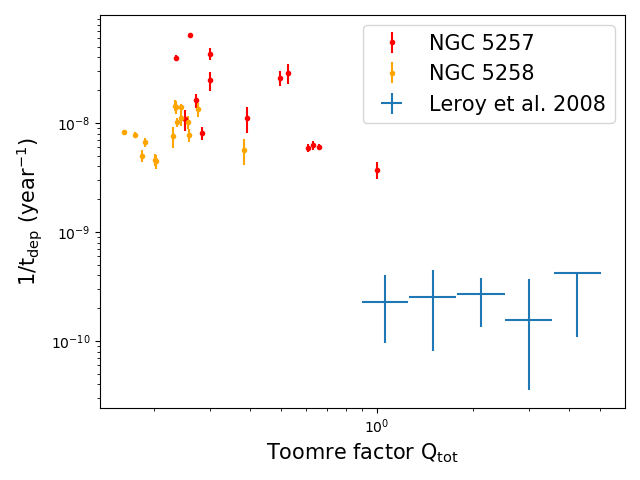}}%
	\subfloat{\includegraphics[width=0.5\linewidth]{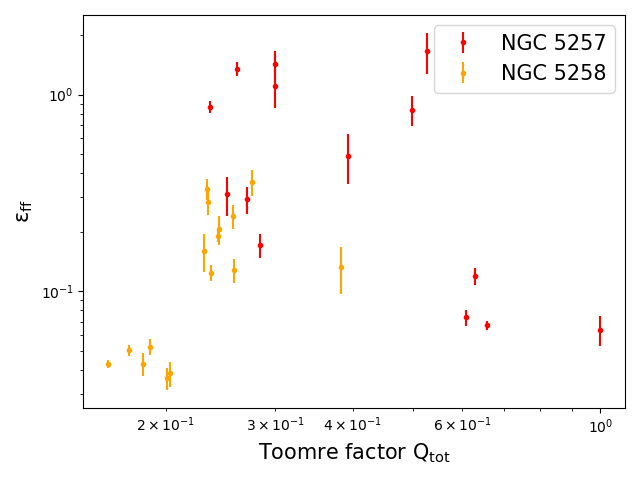}}%
	\caption{(Left) Inverse of depletion time as a function of the Toomre factor. The blue points are the mean values of spiral galaxies from \citet{Leroy_2008}. (Right) SFE per free-fall time as a function of Toomre factor. In both plots $Q_{\mathrm{tot}}$ includes both gas and stellar components. Note that all the points in NGC 5258 arise from the south arm. }
	\label{fig:SFE_Q}
\end{figure*}

We also plotted $t_{\text{dep}}$ and $\epsilon_{\text{ff}}$ versus $Q_{\mathrm{tot}}$ for the star forming pixels, as shown in Fig. \ref{fig:SFE_Q}. We see there is no clear trend in both plots. This is consistent with \citet{Leroy_2008}, who conclude that the Toomre factor only sets the criterion for collapse but does not determine how fast clouds form. This plot further excludes another factor that could drive the high $\epsilon_{\text{ff}}$ in Section 5.1. As we can see in the right plot of Fig. \ref{fig:SFE_Q}, the south arm in NGC 5258 has the lowest $Q_{\mathrm{tot}}$ value among all the star-forming regions but does not show the problem of extreme $\epsilon_{\text{ff}}$ values. 
	
Finally, we note that the Toomre instability analysis assumes the galaxy disk to be infinitely thin without any external force. Therefore, this kind of analysis might not be ideal for mergers.

\subsection{Causes to the high $\mathbf{\epsilon_{\text{ff}}}$}

In Section 5.1, we found many star forming regions in NGC 5257 have $\epsilon_{\text{ff}}$ close to or greater than 100\%. As discussed in Section 5.1, this extreme value is not driven by the underestimate of the stellar potential. We further excluded the factor of Toomre instability in Section 5.2, as the south arm of NGC 5258, which shows similarly low Toomre factor, does not show extreme values of $\epsilon_{\text{ff}}$. The other factors we need to consider are the \cotwo/1-0 ratio value and the \alphaco\ we adopted to calculate $\Sigma_{\text{mol}}$. As shown in Section 3.2, the \cotwo/1-0 line ratio is generally uniform across both galaxies with the exception of the south region in NGC 5257, which has a ratio a factor of 1.6 higher than the global ratio. According to equation \ref{equ: epsff1}, $\epsilon_{\text{ff}}$ will drop by a factor of 2.6 if we adopt this local ratio for that region. However, changing the \cotwo/1-0 ratio does not help for the other two off-center star forming regions in NGC 5257. The \alphaco\ value is highly uncertain and can vary from region to region, which makes the comparison of $t_{\text{dep}}$ and $\epsilon_{\text{ff}}$ among different regions quite difficult. However, according to our RADEX modeling, all of our modeled regions have \alphaco\ value close to the commonly used ULIRG value \citep{Downes_Solomon_1998}. As we have discussed in Section 4.4, the real \alphaco\ value for these regions is likely to be even lower. If this is the case, we will overestimate the $\Sigma_{\text{mol}}$ and thus underestimate the $\epsilon_{\text{ff}}$, which would help the problem even worse. On the other hand, we should note that pixels with extreme $\epsilon_{\text{ff}}$ value are generally from regions with high $\Sigma_{\text{SFR}}$ but relatively low $\Sigma_{\text{mol}}$. Molecular gas in these regions tends to be hotter and more diffuse, which should result in low \alphaco. Thus, \alphaco\ is also unlikely the reason we overestimate the $\epsilon_{\text{ff}}$.  	 
	
As we have mentioned, those pixels with high $\epsilon_{\text{ff}}$ are generally from the off-center star forming regions in NGC 5257 which contain YMCs. If most of the 33 GHz emission in these regions comes from these YMCs, which can be tested using $\sim$pc resolution radio continuum data, we may overestimate $\epsilon_{\text{ff}}$ in the following two aspects. First, we should note that YMCs generally reside in dense cores, which have sizes of $\sim$ pc and gas densities of $\sim$ 10$^4$ cm$^{-3}$ \citep{Leroy_2018}. On the one hand, our \cotwo\ resolution ($\sim$ 500 pc) is not high enough to resolve the local dense core structure. As a result, the local surface density of those regions will be underestimated due to the beam smearing effect. On the other hand, the local gas density is better traced by dense gas tracers such as CN and HCN. The volume density we get from \cotwo\ data should trace the mean volume density along the line of sight, which will also underestimate the local density. In both cases, we will overestimate the free-fall time and thus overestimate $\epsilon_{\text{ff}}$. To resolve this issue, we need CN or HCN data with $\sim$pc resolution. Second, according to \citet{Longmore_2014}, YMCs are generally formed within 1 Myr. As all of these clusters have ages greater than 1 Myr, we would expect the star forming process has finished and the stellar feedback is dominant. Therefore, the radio continuum is tracing the bulk mass of the YMC instead of the SFR and the $\epsilon_{\text{ff}}$ we measured may not make physical sense. To resolve this issue, we can try to characterize $\epsilon_{\text{ff}}$ at the cluster forming stage. In this case, SFR is calculated as $M_{\star}$ traced by radio continuum divided by characteristic cluster form timescale, which is generally set to be 0.5 Myr \citep{Khullar_2019}. However, we need to note that these YMCs are probably in stage 4 according to the classification paradigm in \citet{Whitmore_2016}, where they have both radio and optical counterpart. At this stage, a large fraction of gas is repelled and therefore we might not get the real gas density while cluster is forming. The overall explanation above is supported by the lack of YMCs in other two regions where $\epsilon_{\text{ff}}$ seems to be more normal. However, we should note that these two regions have high $\Sigma_{\text{mol}}$ and therefore any YMCs in these regions would probably be obscured by dust in optical wavelength. High resolution radio continuum data can also help us test if there are hidden YMCs in these regions.

On the other hand, we should note that even though we have all the corrections made above, we might still get $\epsilon_{\text{ff}}$ $\sim100\%$. According to the simulation in \citet{Khullar_2019}, star clusters formed in extremely dense regions ($\Sigma_{\text{mol}} > 10^3\  \si{M_{\odot}\ pc^{-2}}$) can have $\epsilon_{\text{ff}}$ $\sim$ 100\%. Although we mainly find high $\epsilon_{\text{ff}}$ in low gas surface density regions, beam smearing fact may lead us to underestimate the gas surface density in those regions.

\section{Conclusions}


In this paper, we have presented new, multi-transition CO data for the early stage merger Arp 240. We have combined these data with previously
published CO, infrared, and radio continuum data to investigate the gas and star formation properties of this interesting merger system. 
Our main conclusions are summarized below.
\begin{itemize}
	\item As a LIRG, Arp 240 has a gas fraction close to typical normal spiral galaxies (assuming a ULIRG conversion factor). The global gas fraction is 0.05 for NGC 5257 and 0.07 for NGC 5258. On the other hand, the global gas depletion time is about 10$^8$ years, which is an order of magnitude lower than typical normal spiral galaxies. Recent studies \citep{Violino_2018, Pan_2018} find that early mergers typically have higher gas fractions but similar SFEs when compared to normal spiral galaxies. As confirmed in simulations, the two galaxies in Arp 240 have been through first passage. So in its present state, Arp 240 might have properties closer to the late stage mergers identified in \citet{Pan_2018}. 
	\item The molecular gas is concentrated in the center of NGC 5257. This is consistent with the theory that gas will inflow towards the center. However, NGC 5258 shows gas concentration in the south spiral arm instead of the center, which is probably caused by tidal effects. 	
	\item Most regions in both galaxies have \co/\tcoone\ and \cotwo/1-0 ratio similar to normal spiral galaxies. NGC 5257 shows an extremely high \co/\tcoone\ ratio in the south west, which is probably caused by the inflow of pristine gas from the outer disk regions. On the other hand, the low \co/\tcoone\ ratio in the center of NGC 5258 suggests a low \abun\ abundance ratio, which might be either caused by overproduction of \tco\ or overconsumption of \co. Some regions in both galaxies have \cotwo/1-0 ratios higher than 1.0, which suggests these regions are not in thermal equilibrium. 
	\item We use RADEX modeling to study the gas properties in \cothree\ detected regions. For the center of NGC 5257 and the south arm of NGC 5258, where \coone\ emission peaks, the molecular gas has relatively low temperature (10 -- 100 K) and high density (10$^4$ -- 10$^5$ cm$^{-3}$). In contrast, the off-center \cothree\ emission peak in NGC 5257 has relatively high temperature (10 -- 1000 K) and low density (10$^3$ -- 10$^4$ cm$^{-3}$), which indicates this region mainly has warm-phase gas. This is consistent with \coone\ and \cotwo\ observation of that region where there is no emission peak there.    
	\item RADEX modeling shows all of the three \cothree\ detected regions in Arp 240 have an \alphaco\ close to the typical ULIRG value. We also performed an LTE analysis across the entire system and calculated the mean \alphaco\ value for both galaxies. The mean \alphaco\ values for both galaxies are also comparable to the typical ULIRG value. However, the calculated \alphaco\ is dependent on the assumed \abun\ abundance ratio. For similar early merger systems like Arp 55, the \abun\ is about $15 \sim 30$ \citep{Sliwa_2017}, which means the conversion factor (0.2 -- 0.5 \alphacou) in Arp 240 may be much lower even than the typical ULIRG conversion factor. Our input [\co]/[\htwo] could also lead to an overestimate of \alphaco. 
	\item We use the 33 GHz image to measure the SFR in different regions. Due to the limited sensitivity, we can only detect individual starburst regions. The off-center starburst regions in NGC 5257, such as the south bright source in NGC 5257, have very short depletion times ($\sim 10^7$ years) comparable to typical ULIRG depletion time ($\sim$ 20 Myr). These regions are generally associated with the occurrence of young massive clusters. However, we need to note that there is a spatial offset between the radio continuum peak and identified clusters using optical data \citep{Linden_2017}. We suspect this offset may be due to the inaccurate coordinate registration of HST data.
	\item We made a map of the two component Toomre factor $Q_{\mathrm{tot}}$ for both galaxies. All of the major star forming regions have $Q_{\mathrm{tot}}$ smaller than 1.0, with most of them located at a local minimum in $Q_{\mathrm{tot}}$. We also explored the correlation of $Q_{\mathrm{tot}}$ with $t_{\text{dep}}$ and $\epsilon_{\text{ff}}$. We found there is no correlation between those quantities, which is consistent with the finding of \citet{Leroy_2008} for normal spiral galaxies. We therefore agree with \citet{Leroy_2008} that $Q_{\mathrm{tot}}$ only set the criterion on gas collapse but does not affect how fast the gas collapses.
	\item We calculated the SFE per free-fall time of different regions with 33 GHz detections as well as individual pixels in these regions. For individual pixels, the SFE per free-fall time can be above 100\%. 
	Those pixels are also from the off-center star forming regions of NGC 5257, which accompany YMCs. Among all of the factors we have discussed, it seems that the most probable reason for the high $\epsilon_{\text{ff}}$ is that the radio continuum emission is dominated by that from the YMCs in those regions and therefore we overestimate the $\epsilon_{\text{ff}}$. To resolve this issue, we need $\sim$pc resolution data of radio continuum, molecular gas and dense gas tracer. On the other hand, \citet{Khullar_2019} show that some YMCs can have $\epsilon_{\text{ff}}$ $\sim$ 100\% in dense gas region. Therefore, we may still measure $\epsilon_{\text{ff}}$ $\sim$ 100\% for those YMCs in their cluster forming stage after we remove all the observational bias.

\end{itemize}

\section*{Acknowledgements}
We thank Sean Linden for sharing the 33 GHz data and the referee for helpful comments that improved the paper. This paper makes use of the following ALMA data: \\
ADS/JAO.ALMA \#2015.1.00804.S. ALMA is a partnership of ESO (representing its member states), NSF (USA) and NINS (Japan), together with NRC (Canada), MOST and ASIAA (Taiwan), and KASI (Republic of Korea), in cooperation with the Republic of Chile. The Joint ALMA Observatory is operated by ESO, AUI/NRAO and NAOJ. The National Radio Astronomy Observatory is a facility of the National Science Foundation operated under cooperative agreement by Associated Universities, Inc. This work is based [in part] on observations made with the \textit{Spitzer} Space Telescope, which is operated by the Jet Propulsion Laboratory, California Institute of Technology under a contract with NASA.

\section*{Data Availability}
This research is based on public data available in the references. The whole dataset can be obtained from the authors
by request.




\bibliographystyle{mnras}
\bibliography{references} 



\appendix

\section{SFE per free-fall time including stellar component}
\label{epsff_star}
\begin{figure*}
	\subfloat{\includegraphics[width=0.5\linewidth]{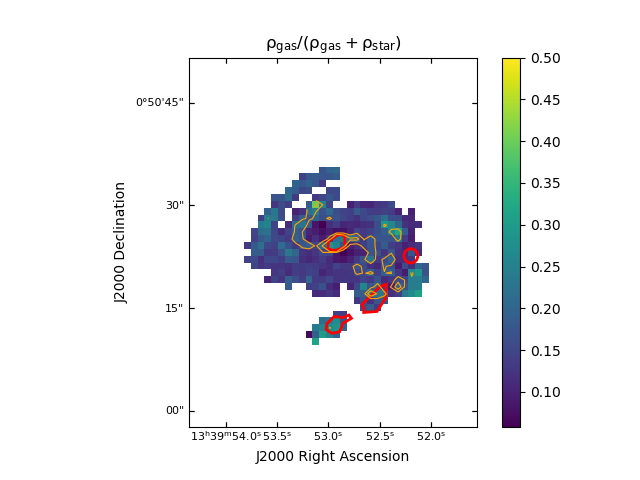}}%
	\subfloat{\includegraphics[width=0.5\linewidth]{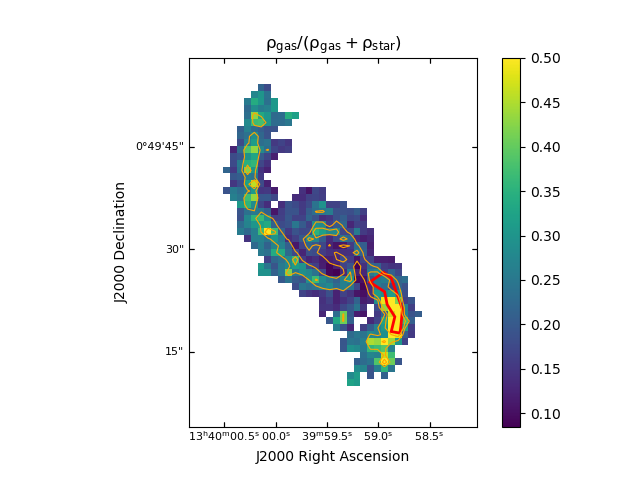}} \\ 
	\caption{Midplane molecular gas fraction for NGC 5257 (left) and NGC 5258 (right). The contours are \cotwo\ moment 0 map (Fig. \ref{fig:mom0}) with levels of 1.1 and 2.2 Jy beam$^{-1}$ km s$^{-1}$. The red polygons are the same as in Fig. \ref{fig:33GHz}.}
	\label{fig:fraction_mid}
\end{figure*}
To calculate the midplane gas fraction, we need to calculate the midplane volume density of stars, which is:
\begin{equation}
\rho_{\star, \mathrm{mid}}=\frac{\Sigma_{\star}}{2H_{\star}}
\end{equation}
where $H_{\star}$ is the scale height of the stellar disk. As we do not have a direct measurement of this quantity, we adopt the empirical equation from \citet{Leroy_2008}, which is
\begin{equation}
H_{\star}=\frac{l}{7.3}
\end{equation}
where $l$ is the scale length of the two galaxies. The stellar component can affect our calculation of the scale height of the disk. Therefore, we need to recalculate the gas scale height to determine the midplane gas volume density. The complete equation for calculating the scale height of the molecular gas is \citep{Wilson_2019}
\begin{equation}
H^{*}_{\mathrm{mol}}=\frac{\sigma_v^2}{\pi G \Sigma_{\text{mol}}}\times \left(\frac{1+\alpha+\beta}{1+\bar{g}_{\mathrm{galaxy}}/\bar{g}}\right) \times \left(\frac{\Sigma_{\text{mol}}}{\Sigma_{\mathrm{tot, GL}}}\right)
\end{equation}
where $\alpha \sim 0.3$ is the magnetic to turbulent plus thermal support ratio and $\beta \sim 0$ is the cosmic ray to turbulent plus thermal support ratio. $\bar{g}_{\mathrm{galaxy}}/\bar{g}$ is the ratio of the average gravitational acceleration of gas caused by the central mass and the galactic disk. This term has a value of 0.05 -- 0.2. Therefore, the second term in the above equation is $\sim 1$. $\Sigma_{tot, GL}$ is the mass surface density of the disk within the gas scale height. So the ratio $\Sigma_{\text{mol}}/\Sigma_{tot, GL}$ can be approximated as the ratio of the midplane density of the gas and the total midplane density. Therefore, the above equation can be simplified as 
\begin{equation}
\label{eqn:scaleheight1}
H^*_{\mathrm{mol}}= \frac{\sigma_v^2}{\pi G \Sigma_{\text{mol}}} \left(\frac{\Sigma_{\text{mol}}}{\Sigma_{\mathrm{tot, GL}}}\right) = \frac{\sigma_v^2}{\pi G \Sigma_{\text{mol}}} \times \frac{\rho^{*}_{\mathrm{mol, mid}}}{\rho^{*}_{\mathrm{mol, mid}}+\rho_{\star, \mathrm{mid}}}
\end{equation}
where $\rho^{*}_{\mathrm{mol, mid}}$ and $\rho_{\star, \mathrm{mid}}$ are the midplane density of molecular gas disk and stellar disk respectively. On the other hand, $\rho^{*}_{\mathrm{mol, mid}}$ can be expressed as
\begin{equation}
\label{eqn:scaleheight2}
\rho^{*}_{\mathrm{mol, mid}}= \frac{\Sigma_{\text{mol}}}{2H^{*}_{\mathrm{mol}}}
\end{equation}
Combining equation \ref{eqn:scaleheight1} and \ref{eqn:scaleheight2}, we explicitly express $\rho_{\mathrm{mol, mid}}^*$ as 
\begin{equation}
\rho^{*}_{\mathrm{mol, mid}}=\frac{\Sigma_{\text{mol}}}{4H_{\mathrm{ch}}}+\sqrt{\frac{\Sigma_{\text{mol}}^2}{16 H_{\mathrm{ch}} ^2}+\frac{\Sigma_{\text{mol}}}{2H_{\mathrm{ch}}} \rho_{\star, \mathrm{mid}}}
\end{equation}
where $H_{\mathrm{ch}}=\frac{\sigma_v^2}{\pi G \Sigma_{\text{mol}}}$ is the characteristic scale height of the molecular gas. We can then calculate the midplane gas fraction 
\begin{equation}
f_{\mathrm{mid}}=\frac{\rho^*_{\mathrm{mol, mid}}}{\rho_{\mathrm{tot, mid}}} = \frac{\rho^*_{\mathrm{mol, mid}}}{\rho_{\star, \mathrm{mid}}+\rho^*_{\mathrm{mol, mid}}}
\end{equation}
The midplane gas fraction maps for both galaxies are shown in Figure \ref{fig:fraction_mid}. 

In our original analysis, we assumed the midplane gas fraction to be 0.5. Therefore, the relationship between $H^{*}_{\mathrm{mol}}$ and $H_{\mathrm{mol}}$ is
\begin{equation}
H^{*}_{\mathrm{mol}}= H_{\mathrm{mol}} \times \frac{f_{\mathrm{mid}}}{0.5}
\end{equation}
Then the relation between $\rho^{*}_{\mathrm{mol, mid}}$ and $\rho_{\mathrm{mol, mid}}$ is 
\begin{equation}
\rho^{*}_{\mathrm{mol, mid}}= \rho_{\mathrm{mol, mid}} \frac{0.5}{f_{\mathrm{mid}}}
\end{equation}
Therefore, the modified free-fall time becomes 
\begin{equation}
t_{\mathrm{ff, mod}}= \sqrt{\frac{3\pi}{32 G \rho^*_{\mathrm{mol, mid}}}}=t_{\text{ff}}\sqrt{\frac{f_{\mathrm{mid}}}{0.5}}
\end{equation}
and the modified efficiency per free-fall time is 
\begin{equation}
\epsilon_{\mathrm{ff, mod}}= \epsilon_{\text{ff}}\sqrt{\frac{f_{\mathrm{mid}}}{0.5}}
\end{equation}

\end{document}